\documentclass[a4paper,11pt]{article}
\pdfoutput=1 

\usepackage{jheppub} 

\usepackage[T1]{fontenc} 
\usepackage{natbib}
\usepackage{graphicx}
\usepackage{amsfonts}
\usepackage{amssymb}
\usepackage{amsbsy}
\usepackage{amsmath}
\usepackage{mathrsfs}
\usepackage{latexsym}
\usepackage{natbib}
\usepackage{bm}
\usepackage{color}
\usepackage{wasysym}
\usepackage{hyperref}
\usepackage{physics}
\usepackage{caption}
\usepackage{subcaption}
\usepackage{float}
\usepackage{soul}

\title{\boldmath The fate of a Quantum-Corrected Collapsing Star in General Relativity}

\author[a,1]{Shibendu Gupta Choudhury\note{Corresponding author},}
\author[b]{Soumya Chakrabarti}

\affiliation[a]{S. N. Bose National Centre for Basic Sciences, \\
JD Block, Sector-III, Salt Lake City, \\
Kolkata - 700 106, India}
\affiliation[b]{Department of Physics\\ School of Advanced Science, \\
Vellore Institute of Technology, Vellore, \\
Tiruvalam Rd, Katpadi, Tamil Nadu 632014, India.}

\emailAdd{shibendu17@bose.res.in}
\emailAdd{soumya.chakrabarti@vit.ac.in}

\abstract{We incorporate some corrections inspired by loop quantum gravity into the concept of gravitational collapse and propose a complete model of the dynamic process. The model carries the essence of a mass-independent upper bound on the curvature scalars  originally found as a crucial feature of black holes in loop quantum gravity. The quantum-inspired interior is immersed in a geometry filled with null radiation and they are matched at a distinct boundary hypersurface. The ultimate fate of the process depends on inhomogeneities of the metric tensor cofficients. We find a critical parameter $\lambda$ embedded in the inhomogeneity of the conformal factor of the interior metric. Examples with $\lambda < 0$ enforce an eventual collapse to singularity and $\lambda > 0$ cases produce a non-singular collapse resulting in a loop-quantum-corrected Schwarzschild geometry modulo a conformal factor. Interestingly, for $\lambda < 0$ as well, there exist situations where the quantum effects are able to cause a bounce but fall short of preventing the ultimate formation of singularity. The trapped surface formation condition is studied for $\lambda<0$ case to infer about the visibility of the final singularity. Interestingly, we find a possibility of formation of three horizons during the course of the collapse. Eventually all of them merge into one single horizon which envelopes the final singularity. For the non-singular case, there is a possibility that the sphere can evolve into a wormhole throat whose radius is found to be inversely proportional to the critical parameter $\lambda$. Depending on the nature of evolution and the shell regions, the collapsing shells violate some standard energy conditions which can be associated with the quantum inspired corrections.}

\begin{document} 
\maketitle
\flushbottom

\section{Introduction}
Among all the fundamental forces, gravity stands out a little based on its foundation principles. The idea being carried forward by the General Theory of Relativity (GR) is that gravity is just a consequence of generalizing the principles of Euclidean geometry into Riemannian. It took about a century of theoretical and half a century of observational expeditions to ensure that this idea can find it's present stature. GR has been through a remarkable journey, yet it has to walk a tightrope on quite a few fronts. The most persisting one is related to black hole singularities. A black hole is a solution of the field equations of GR and hosts two intriguing features : (i) a central singularity, characterized by the divergence of curvature invariants and (ii) the formation of a horizon that hides this divergence from an external observer. The space-time geometry becomes increasingly curved while approaching this geometry. Even a photon emitted by an object near the horizon can take forever to push through the ever-increasing gravitational pull and as a result, light of the object seems frozen at the horizon. Although the singularity is `technically' hidden behind a horizon, it is not wise to simply ignore them. Using a Penrose diagram it is possible to argue that any central singularity may persist as part of the future boundary of space-time even after the host black hole is completely evaporated. A quantum gravity proposition to resolve this singularity issue is expected for quite some time now, however, no concrete resolution has been found from popular formalisms of quantum gravity, atleast not yet. This motivates a relativist to study natural laws around an $r \simeq 0$ singularity with more rigor and try to reveal any information on \textit{physics beyond Einstein}. \\

One has to keep in mind that this problem has more than one paths of approach. Moreover, one must visualize the entire process of black-hole formation as a dynamical evolution starting from a massive stellar distribution. This process can be identified as a gravitational \textit{collapse} \cite{Datt1938, PhysRev.55.374, PhysRev.56.455}. The name originates from the fact that an initially massive distribution is driven into an indefinite \textit{collapse} by an overpowering gravitational pull towards the centre (see for instance \cite{Malafarina2016, Joshi2000} for a short summary). A singularity or an infinitely dense point can then be classified as a black hole if and only if there is a formation of null surface shielding any exchange of information (matter/radiation). By exploring the dynamical process one may also come across feasible constructions known as an exposed/naked singularity \cite{Yodzis1973, MllerzumHagen1974, PhysRevD.65.101501, PhysRevD.70.087502}. This can happen either due to an absence of null surface \cite{doi:10.1142/S0217732300000992} or a delayed formation due to some extreme physical conditions (e.g., generation of shock waves due to shear \cite{PhysRevD.65.101501} or departure from spherical symmetry \cite{bronni}). A compromise can be designed in the form of a \textit{Cosmic Censorship Hypothesis} \cite{Penrose2002} and argue that any formation of singularity is necessarily accompanied by a horizon. However, there has never been any concrete proof of such a conjecture, and the counter-examples are aplenty.    \\ 

A diplomatic resolution to this impasse is to look for models of gravitational collapse with no singularities. There are \textit{few} examples of this sort, where mostly, the non-singular end-state finds its basic motivation from exotic matter constituents \cite{chakra}. Models involving a \textit{Critical Phenomena} \cite{PhysRevLett.70.9} deserves particular mention, where the outcome of a collapse depends on parameters characterising the initial data. For specific choices of these parameters, an initially collapsing scalar field disperses away leaving behind an empty flat space \cite{Gundlach2007}. Gravitational collapse leading to a non-singular Lorentzian wormhole geometry has also been reported recently \cite{PhysRevD.104.024071}. Very limited examples also exist in modified theories such as torsion-based \cite{PhysRevD.101.124044} and higher curvature theories of gravity \cite{Chakrabarti2018, Chakrabarti2018egb}. They do not necessarily point out any general law of a non-singular collapse but rather serve as specific examples. We try to motivate the same from a different intuition. If a gravitational collapse leads a general relativistic geometry towards zero length scale, it is more logical to expect a \textit{quantum corrected Schwarzschild black hole} as an outcome, opposed to its classical counterpart. Moreover, if we are to seek an effective quantum description of black holes, we must consider that a quantum signature should start appearing gradually, not as an instantaneous effect, even on a scale much larger than the Planck length scale. During this process we should expect the system to encounter a \textit{critical/cross-over} phase. This phase involves an interior geometry that has already reached the quantum corrected form and a classical exterior, written as a vacuum or radiation-dominated solution of GR. The boundary hypersurface separating these two geometries is our point of interest. It is crucial to define and match the tensorial fundamental forms, namely, metric and extrinsic curvature of the two metrics (interior and exterior) near this hypersurface \cite{santosbm, Kolassis1988, KOLASSIS1989243, oliveira, PhysRevD.45.2732, doi:10.1142/S0218271805006584}. We discuss that correcting the interior geometry into a quantum corrected geometry leads to a few non-trivial modifications if we demand a smooth matching at the boundary \cite{PhysRevLett.96.031103, PhysRevD.94.104056, DeLorenzo2015, PhysRevD.96.044010, PhysRevD.98.104016}.    \\

The metric we propose can be regarded as a time-evolving analogue of a recently proposed quantum-corrected Schwarzschild solution \cite{PhysRevLett.121.241301, PhysRevD.98.126003, doi:10.1142/S0218271820500765}. The pros and cons of including a quantum correction perturbatively in a classical metric tensor has been discussed in literature at some length \cite{PhysRevD.76.104030, PhysRevD.78.064040, Corichi_2016, Olmedo_2017, BOUHMADILOPEZ2020100701, PhysRevLett.121.241301, PhysRevD.98.126003, doi:10.1142/S0218271820500765}. We focus entirely on the time-evolution of such quantum effects in the physical process involving the death of a spherical star. Our proposed solution is a metric conformally related to the quantum corrected static solution as in \cite{doi:10.1142/S0218271820500765}. Executing the boundary matching we prove that for a set of reasonable initial conditions, the collapse should lead to non-singular outcomes. We also discuss an idea that during the collapse the geometry can sometime behave like the throat of a wormhole. From a technical point of view, a novel feature of this work is an extensive use of extrinsic curvature matching equation \cite{santosbm, chanbm}, while the field equations are used only as constraints of the energy momentum components.

\section{The Quantum Corrected Static Geometry}
The most crucial feature which the recently proposed quantum-corrected black hole geometry \cite{PhysRevLett.121.241301, PhysRevD.98.126003, doi:10.1142/S0218271820500765} carries is a mass-independent upper bound for the curvature scalars. The solution stands out due to the parameter $\epsilon \sim m^{-\frac{2}{3}}$, which signifies loop-quantum corrections with an estimated value of $\epsilon \sim 10^{-26}$ for solar mass black holes \cite{sym12081264}. One can write the static metric with first order corrections as \cite{doi:10.1142/S0218271820500765},
\begin{equation}\label{qcorsh}
\begin{split}
 ds^2=-\left( \frac{r}{r_s}\right)^{2\epsilon}\left[ 1-
\left(\frac{r_s}{r}\right)^{1+\epsilon}  \right] dt^2+
\frac{dr^2}{ 1-
 \left(\frac{r_s}{r}\right)^{1+\epsilon}}  +r^2 \left(d\theta^2+\sin^2\theta d\phi^2\right),
 \end{split}
\end{equation}
where $r_s = 2m$, $m$ being the mass of the black hole. In a limit of $\epsilon \rightarrow 0$, one should get back the Schwarzschild solution of GR. Note that the effective energy-density outside the horizon is given by \cite{doi:10.1142/S0218271820500765},
\begin{equation}
\rho = - \frac{\epsilon}{8\pi r^2} \left( \frac{r_s}{r}
\right)^{1+\epsilon},
\end{equation}
which is negative.
 
\section{Construction of a Time-evolving Analogue}
We construct a time-evolving analogue metric of Eq. \eqref{qcorsh} which can describe the gravitational collapse of a quantum corrected spherical interior,
\begin{eqnarray}\label{intmet}
&& ds_-^2=\left[A(t)+B(r)\right]^2\left[-f^2(r)dt^2+g^2(r)dr^2\right. \left. +r^2 d\theta^2+r^2 \sin^2\theta d\phi^2\right], \\&&
f^2(r)=\left( \frac{r}{r_s(r)}\right)^{2\epsilon(r)}\left[ 1- \left(\frac{r_s(r)}{r}\right)^{1+\epsilon(r)} \right],\\&&
g^2(r)=\left[1-\left(\frac{r_s(r)}{r}\right)^{1+\epsilon(r)}\right]^{-1}.
\end{eqnarray}

Evidently, the time evolution comes into the metric as an inhomogeneous conformal factor $[A(t)+B(r)]^2$. We further assume the collapsing fluid to be imperfect and write the corresponding energy momentum tensor as,
\begin{equation}\label{emtensor}
\begin{split}
 \left(T_{\alpha\beta}\right)
=\left(\rho+p_\mathrm{t}\right)u_\alpha u_\beta+p_\mathrm{t} g_{\alpha\beta}
+\left(p_\mathrm{r}-p_t\right)\chi_\alpha \chi_\beta+ q\left(u_\alpha \chi_\beta+ u_\beta \chi_\alpha\right).
\end{split}
\end{equation}
$\rho$, $p_\mathrm{r}$, $p_\mathrm{t}$ and $q$ are the energy density, radial pressure, tangential pressure and radial heat flux of the fluid, respectively. $\chi^\alpha$ is a unit vector along the radial direction and $u^\alpha$ is the four-velocity of the fluid, given by,
\begin{equation}\label{uvs}
 u^\alpha=\frac{1}{\left[A(t)+B(r)\right]f(r)}\delta^\alpha_0, \hspace{0.2cm} \chi^\alpha = \frac{1}{\left[A(t)+B(r)\right]g(r)}\delta^\alpha_1,
\end{equation}
where $u^\alpha u_\alpha = -1$, $\chi^\alpha \chi_\alpha = 1$ and $u^\alpha\chi_\alpha=0$. With this we can write the Einstein field equations as,
\begin{equation}\label{fe1}
\begin{split}
 \frac{1}{r^2 g^3 (A+B)^4 f^2}&\left.\Bigg[ r^2 g \left(3 g^2 \dot{A}^2+f^2 B'^2\right)+A^2 f^2 \left(2 r g'+g^3-g\right) \right. \\ & \left. +2 A f^2  \left\lbrace r \left(B' \left(r g'-2 g\right)-r g B''\right)+B \left(2 r g'+g^3-g\right)\right\rbrace \right. \\ & \left. -2 r B f^2 \left\lbrace r g B''+B' \left(2 g-r g'\right)\right\rbrace+B^2 f^2 \left(2 r g'+g^3-g\right)\right.\Bigg] =\kappa\rho,
 \end{split}
 \end{equation}

\begin{equation}\label{fe2}
 \frac{2 \dot{A} \Big[f'(A+B)+2 f B'\Big]}{f^2 g (A+B)^4}=-\kappa q,
 \end{equation}

\begin{equation}\label{fe3}
\begin{split}
\frac{1}{r^2 g^2 f^2 (A+B)^4} & \Bigg[\left.2 r B\left\lbrace f B' \left(r f'+2\right)-r g^2 \ddot{A}\right\rbrace  +r^2 \left(g^2 \dot{A}^2+3 f^2 B'^2\right) \right. \\ & \left. +2 A \left\lbrace r \left[f B' \left(r f'+2 f\right)-r g^2 \ddot{A}\right]+B f \left(2 r f'-f g^2+f\right)\right \rbrace \right. \\ & \left. +A^2 f \left(2 r f'-f g^2+f\right)+B^2 f \left(2 r f'-f g^2+f\right)\right.\Bigg]=\kappa p_\mathrm{r},
 \end{split}
 \end{equation}
and
\begin{equation}\label{fe4}
\begin{split}
 \frac{1}{r f^2 g^3 (A+B)^4} & \left. \Bigg[ r g^3 \left\lbrace A'^2-2 A'' (A+B)\right\rbrace-f g' (A+B) \left\lbrace f \left(A+2 r B'+B\right)+r f' (A+B)\right\rbrace \right. \\ & \left. +f g \Big\{ A^2 \left(r f''+f'\right) -r f B'^2 +2 A \left[\left(r B'+B\right) f'+f \left(r B''+B'\right)+r B f''\right]\right. \\ &  +B^2 \left(r f''+f'\right) +2 B \Big[r f B''+B' \left(r f'+f\right)\Big]\Big\} \Bigg]=\kappa p_\mathrm{t}.
 \end{split}
\end{equation}

Derivatives with respect to $t$ and $r$ are written using `dot's and `prime's respectively. In the subsequent sections we use $\kappa = 1$.

\subsection{Boundary Matching Conditions}
Since the collapsing interior has a radial heat flux, we consider the exterior geometry to have a Vaidya structure, representing a spherically symmetric null fluid -
\begin{equation} \label{v1}
ds^2_{+} = - \left(1 - \frac{2M(v)}{R}\right) dv^2 - 2du dR + R^{2}\left(d\theta^2+ \sin^2\theta d\phi^2\right).
\end{equation}
$v$ is the retarded time. $M(v)$ is the mass of the spherical distribution enclosed within the boundary surface $\Sigma$. A smooth matching of two fundamental forms, the metric and the extrinsic curvature at the boundary hypersurface is written as the following conditions,
\begin{equation}
ds^2_{-} \overset{\Sigma}{=} ds^2_{+},
\label{eq:dsidso}
\end{equation}
and
\begin{equation}
K^{-}_{ij} \overset{\Sigma}{=} K^{+}_{ij}.
\label{eq:kijikijo}
\end{equation}
The components of extrinsic curvature $K^{\pm}_{ij}$-s are defined as,
\begin{equation}
K^{\pm}_{ij}=-n^{\pm}_{\alpha}
{{\partial^2x^{\alpha}} \over {\partial \xi^i \partial \xi^j}}
-n^{\pm}_{\alpha}\Gamma^{\alpha}_{\beta \gamma}
{{\partial x^{\beta}}  \over {\partial \xi^i}}
{{\partial x^{\gamma}} \over {\partial \xi^j}}.
\label{eq:kij}
\end{equation}

We have written the unit normal vectors to $\Sigma$ as $n^{\pm}_{\alpha}$-s. $x^{\alpha}$-s are coordinates of two spacetimes and the $\xi^i$-s are coordinates on the surface $\Sigma$. $\Gamma^{\alpha}_{\beta \gamma}$-s denote the standard Christoffel symbols. For a detailed discussion on matching of fundamental forms, we refer to \cite{Israel1967, santosbm, 10.1046/j.1365-8711.2000.03547.x} and references therein. If an interior metric carries the general form,
\begin{equation}
\label{met-int}
ds^2 = -S^2(t,r)dt^{2} + N^2(t,r)dr^{2} + C^2(t,r)(d\theta^{2} + \sin^2\theta d\phi^{2}),
\end{equation}
one can derive the total energy enclosed within the surface $\Sigma$ as \cite{10.1046/j.1365-8711.2000.03547.x},
\begin{equation}\label{msmf}
 M\overset{\Sigma}{=}\left[\frac{C}{2}\left(1+\frac{\dot{C}^2}{S^2}-\frac{{C^\prime}^2}{N^2}\right)\right].
\end{equation}
This is also known as the Misner and Sharp mass function \cite{PhysRev.136.B571, 1966ApJhm, 10.1063/1.1665273}.
The extrinsic curvature matching conditions can be simplified to write a differential equation involving the interior metric coefficients as,
\begin{equation}\label{extrinscurv}
2\left(\frac{\dot{C'}}{C}-\frac{\dot{C}S'}{CS}-\frac{\dot{N}C'}{NC}\right)\overset{\Sigma}{=}
-\frac{N}{S}\left[\frac{2\ddot{C}}{C}-\left(\frac{2\dot{S}}{S} -\frac{\dot{C}}{C}\right) \frac{\dot{C}}{C}\right]+\frac{S}{N}\left[\left(\frac{2S'}{S}+\frac{C'}{C}\right)\frac{C'}{C}-\left(\frac{N}{C}\right)^2\right].
\end{equation}

For the metric in Eq. \eqref{intmet} this condition simplifies into,
\begin{equation}\label{Teq}
 \frac{2\ddot{T}}{T}-\left(\frac{\dot{T}}{T}\right)^2-\frac{2\dot{T}}{T}\left(\frac{x_1}{T}+x_2\right)-\frac{1}{T}\left(\frac{x_3}{T}+x_4\right)-x_5=0,
 \end{equation}
where
\begin{equation}
\begin{split}
& T(t)=A(t)+B(r_0),
\hspace{0.2cm}\\ & x_1=\frac{2 B'(r_0) f(r_0)}{g(r_0)}, \hspace{0.2cm} x_2=\frac{f'(r_0)}{g(r_0)}, \hspace{0.2cm} x_3=3\left[\frac{B'(r_0)f(r_0)}{g(r_0)}\right]^2, \\
& x_4=2B'(r_0)\frac{f^2(r_0)}{g^2(r_0)}\left[\frac{2}{r_0}+\frac{f'(r_0)}{f(r_0)}\right], \hspace{0.2cm} x_5=\frac{f^2(r_0)}{g^2(r_0)}\left[\frac{2f'(r_0)}{r_0 f(r_0)}+\frac{1}{{r_0}^2}\right]-\frac{f^2(r_0)}{r_0^2}.
\end{split}
\end{equation}

We have taken the radial coordinate at the boundary as $r = r_0$. Since $r_0$ is simply any arbitrary value, we can simply use this equation to solve for a time evolution of the collapsing system under consideration.

\section{Fate of the Collapse}
We solve Eq. \eqref{Teq} numerically in order to find the behavior of $T(t)$. For this, first we specify the inhomogeneities in the the metric tensor, i.e., the functional forms of $B(r)$, $r_s(r)$ and $\epsilon(r)$,
\begin{equation}\label{inhomo}
 B(r)=\alpha  \exp \left[\lambda  (r_0-r)\right], \hspace{0.2cm} r_s(r)=\beta  r \exp (r-r_0), \hspace{0.2cm} \epsilon (r)=\gamma \left[\exp (r-r_0)\right]^{-\frac{2}{3}},
\end{equation}
where $\alpha$, $\beta$, $\lambda$ and $\gamma$ are constants. $r_0$ is the value of radial coordinate at an arbitrary boundary surface. We can classify the nature of evolution into two cases, corresponding to $\lambda = -1$ and $\lambda = 1$. In the following subsections we consider each of them individually.

\subsection{Case I: $\mathbf{\lambda = -1}$}
We obtain the numerical solution with the initial conditions : $T(0) = 10$ and $\dot{T}(0) = -0.1$. This ensures that we are constructing an initially collapsing spherical distribution with a finite radius of two-sphere. We have chosen $\beta = 0.5$ to ensure $\frac{r_s}{r} < 1$ and $\gamma = 0.001$ to keep $\epsilon$ small.
\subsubsection{Evolution}
\begin{figure}[H]
     \centering
    \boxed{  \begin{subfigure}[b]{0.45\textwidth}
         \centering
         \includegraphics[width=\textwidth]{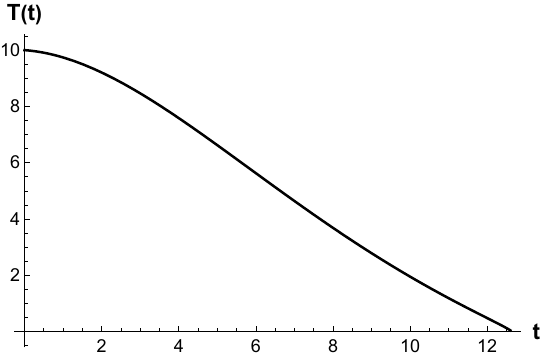}
         \caption{$\alpha=1$}
         \label{fig1a}
     \end{subfigure}
     \hfill
     \begin{subfigure}[b]{0.45\textwidth}
         \centering
         \includegraphics[width=\textwidth]{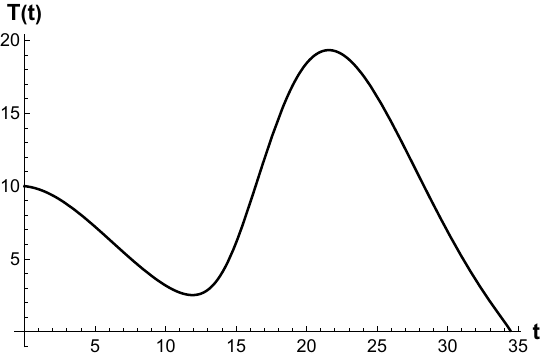}
         \caption{$\alpha=2$}
         \label{fig1b}
     \end{subfigure}
     \hfill}
        \caption{Case I - Evolution of the temporal part of the conformal factor, $T(t)$ with $t$ for different values of $\alpha$ where we have chosen $r_0=4$.}
        \label{fig1}
\end{figure}

For the negative $\lambda$ case the collapse always hits a singularity, irrespective of the values of $\alpha$ and $r_0$. This is evident from the plot of the radius of two-sphere in Fig. \ref{fig1}. In addition we find that the qualitative behavior of this evolution does not change for different values of $r_0$. For $\alpha=1$, $T$ decreases monotonically and reaches zero after a finite time. However, for $\alpha \geq 2$, alternative phases of bounce and implosion can be seen before $T$ finally goes to produce a zero proper volume. The total number of bounce depends on the value of $\alpha$ one chooses. Since there is an inhomogeneity in the conformal factor, the collapsing shells (realized by putting different values of $r \leq r_0$) do not hit singularity simultaneously. The regions near the center hits the singularity prior to the regions near the boundary. For more insights about these shells we study the validity of different energy conditions for different shells, as a function of time.

\subsubsection{Energy Conditions}
\begin{figure}[H]
     \centering
  \boxed{   \begin{subfigure}[b]{0.45\textwidth}
         \centering
         \includegraphics[width=\textwidth]{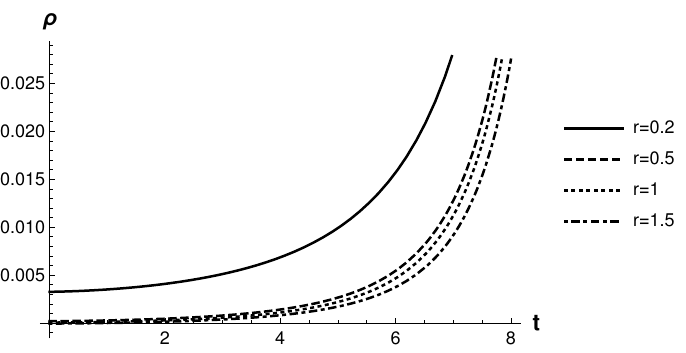}
         \label{fig2a}
     \end{subfigure}
     \hspace{0.2cm}
     \begin{subfigure}[b]{0.45\textwidth}
         \centering
         \includegraphics[width=\textwidth]{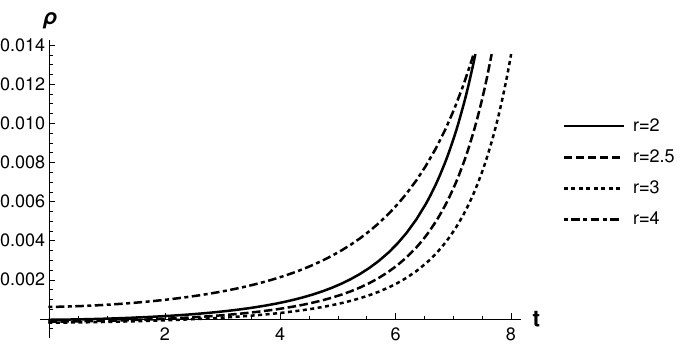}
         \label{fig2b}
     \end{subfigure}
      }
         \caption{Case I - Evolution of the energy density, $\rho$ with $t$ at different regions characterized by different values of $r$, with $r_0=4,\alpha=1$.}
        \label{fig2}
\end{figure}

\begin{figure}[H]
     \centering
    \boxed{ \begin{subfigure}[b]{0.45\textwidth}
         \centering
         \includegraphics[width=\textwidth]{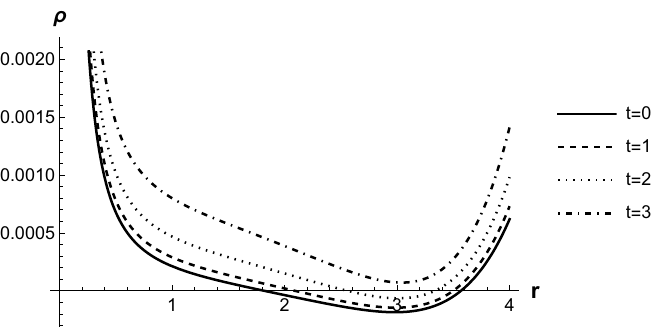}
         \label{fig3a}
     \end{subfigure}
     \hspace{0.2cm}
     \begin{subfigure}[b]{0.45\textwidth}
         \centering
         \includegraphics[width=\textwidth]{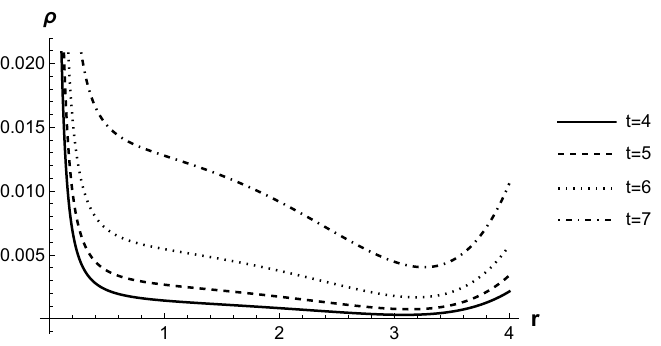}
         \label{fig3b}
     \end{subfigure}}
         \caption{Case I - Plot of the energy density profile $\rho(r)$ for different instants of time $t$, with $r_0=4$, $\alpha=1$.}
        \label{fig3}
        \end{figure}

The Weak Energy Condition (WEC) is given by,
 \begin{equation}
  T_{\mu\nu}u^\mu u^\nu\geq 0 \implies \rho\geq 0,
 \end{equation}
which ensures the non-negativity of the energy density of a classical fluid distribution. Using Eq. \eqref{fe1}, we study the behavior of $\rho(r,t)$. For $\alpha = 1$, the evolution of $\rho(r,t)$ is shown in Figs. \ref{fig2} and \ref{fig3}. In Fig. \ref{fig2}, we have plotted the WEC (energy density) during the evolution for different shells (i.e., for different values of $r \leq r_0$). Fig. \ref{fig3} represents the behavior of WEC as a function of distance from the centre for given values of time $t$. The plots suggest that the WEC is not always satisfied throughout the distribution. In a certain range of $r$ the WEC shows clear violation as soon as the collapse begins. After a finite time, enough matter is drawn towards the collapsing core and as a consequence, the WEC is satisfied. Violation of the WEC can easily be connected to the quantum effects which can give rise to an effective negative energy density of the system, unless the matter energy density is high enough to dominate.

\begin{figure}[H]
     \centering
    \boxed{ \begin{subfigure}[b]{0.45\textwidth}
         \centering
         \includegraphics[width=\textwidth]{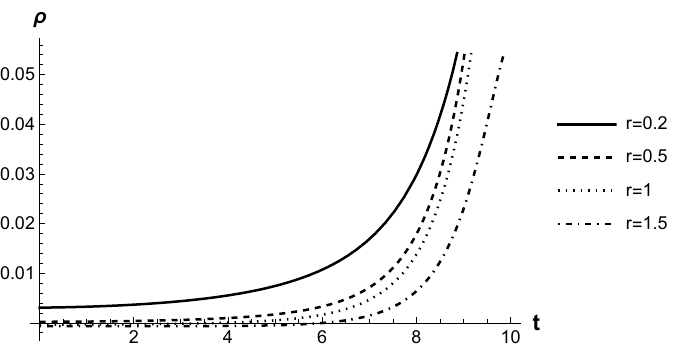}
         \label{fig4a}
     \end{subfigure}
     \hspace{0.2cm}
      \begin{subfigure}[b]{0.45\textwidth}
         \centering
         \includegraphics[width=\textwidth]{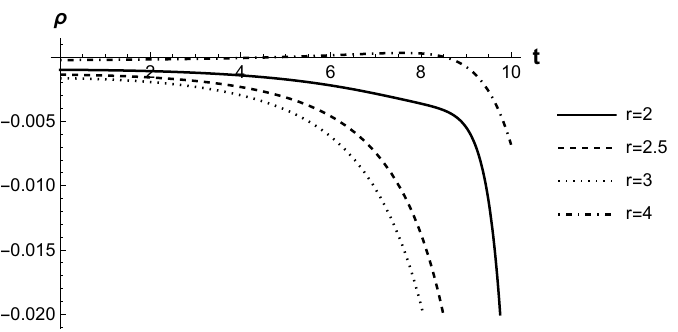}
         \label{fig4b}
     \end{subfigure}}

        \caption{Case I - Evolution of the energy density, $\rho$ with $t$ at different regions characterized by different values of $r$, with $r_0=4,\alpha=2$.}
        \label{fig4}
\end{figure}

\begin{figure}[H]
     \centering
   \boxed{  \begin{subfigure}[b]{0.45\textwidth}
         \centering
         \includegraphics[width=\textwidth]{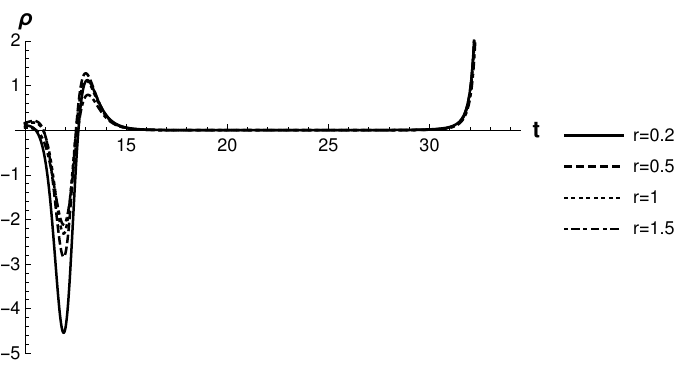}
         \label{fig5a}
     \end{subfigure}
     \hspace{0.2cm}
     \begin{subfigure}[b]{0.45\textwidth}
         \centering
         \includegraphics[width=\textwidth]{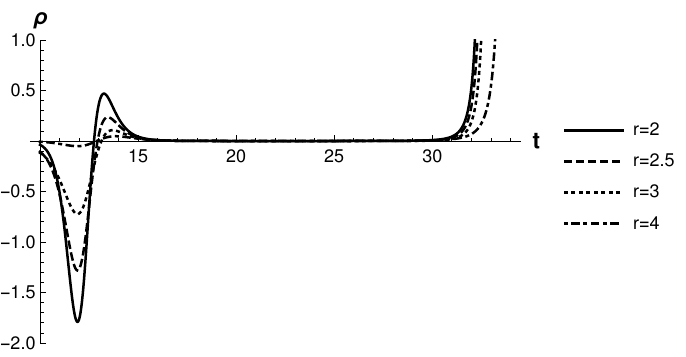}
         \label{fig5b}
     \end{subfigure}
     \hfill}
        \caption{Case I - Evolution of the energy density, $\rho$ with $t$ at different regions characterized by different values of $r$, with $r_0=4,\alpha=2$.}
        \label{fig5}
\end{figure}

\begin{figure}[H]
     \centering
   \boxed{  \begin{subfigure}[b]{0.45\textwidth}
         \centering
         \includegraphics[width=\textwidth]{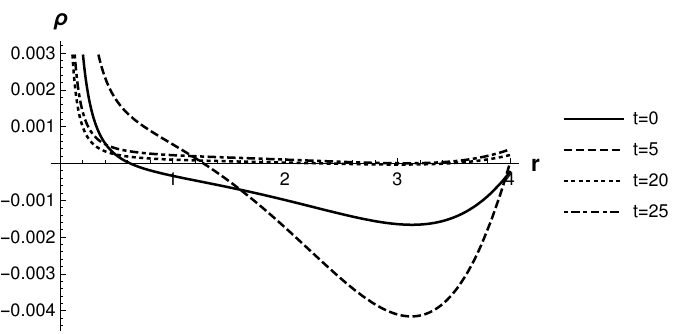}
         \label{fig6a}
     \end{subfigure}
     \hspace{0.2cm}
     \begin{subfigure}[b]{0.45\textwidth}
         \centering
         \includegraphics[width=\textwidth]{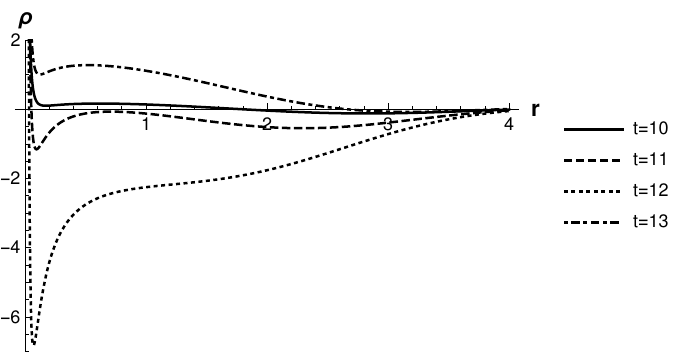}
         \label{fig6b}
     \end{subfigure}
     \hfill}
        \caption{Case I - Plot of the energy density profile $\rho(r)$ for different instants of time $t$, with $r_0=4$, $\alpha=2$.}
        \label{fig6}
\end{figure}

For a higher value of $\alpha$, there are consecutive phases of bounce and collapse and the behavior of WEC is expected to change. We consider the case of $\alpha = 2$ and plot the corresponding evolution of WEC for different shells in Figs. \ref{fig4} and \ref{fig5}. The behavior is shown using two plots for different ranges of time. The radial distribution of the condition for different values of time is shown in Fig. \ref{fig6}. It is quite clear that a bounce in the evolution is always accompanied by a drop (towards negativity!) in the energy density. It is natural to think that a bounce in this model is only possible due to quantum corrections which also generates a negative energy density.
 
\begin{figure}[H]
     \centering
    \boxed{ \begin{subfigure}[b]{0.45\textwidth}
         \centering
         \includegraphics[width=\textwidth]{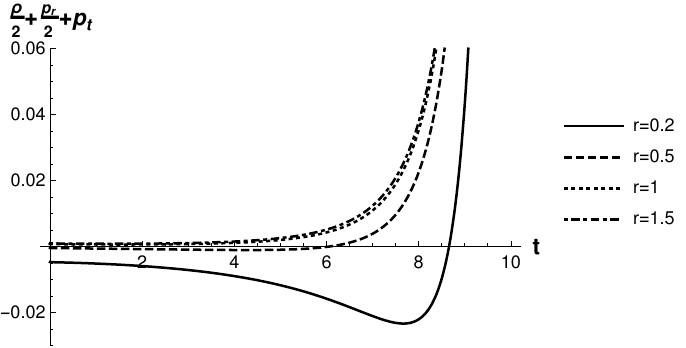}
         \label{fig7a}
     \end{subfigure}
     \hspace{0.2cm}
     \begin{subfigure}[b]{0.45\textwidth}
         \centering
         \includegraphics[width=\textwidth]{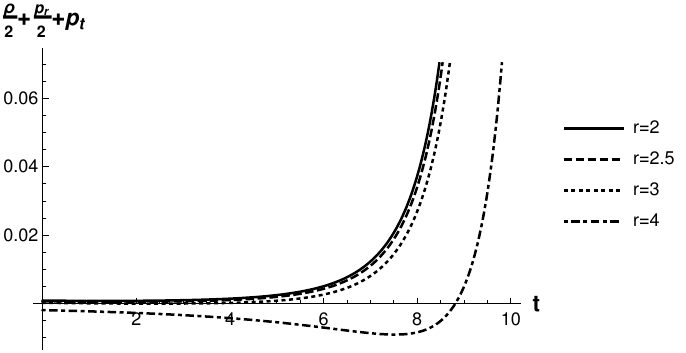}
         \label{fig7b}
     \end{subfigure}}
         \caption{Case I - Study of the SEC during the evolution at different regions characterized by different values of $r$, with $r_0=4,\alpha=1$.}
        \label{fig7}
\end{figure}

\begin{figure}[H]
     \centering
   \boxed{  \begin{subfigure}[b]{0.45\textwidth}
         \centering
         \includegraphics[width=\textwidth]{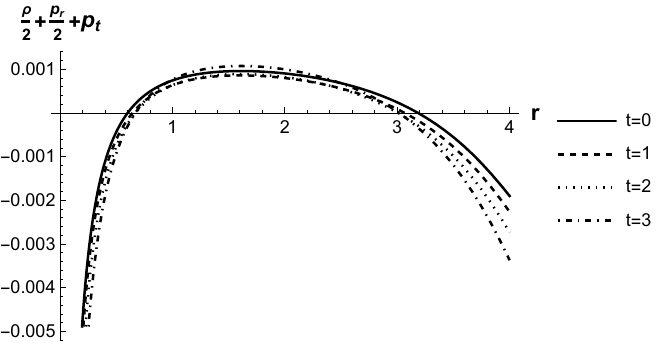}
         \label{fig8a}
     \end{subfigure}
     \hspace{0.2cm}
     \begin{subfigure}[b]{0.45\textwidth}
         \centering
         \includegraphics[width=\textwidth]{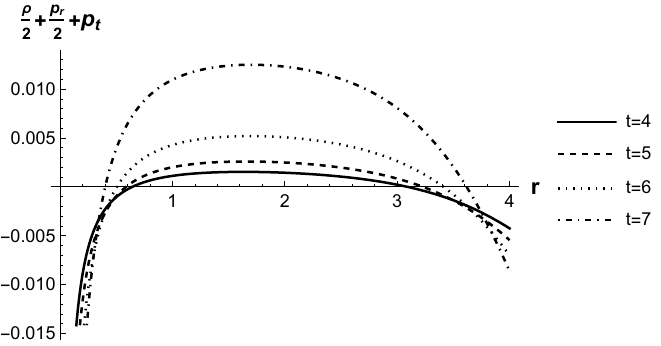}
         \label{fig8b}
     \end{subfigure}}
       \caption{Case I - Study of the SEC within the whole distribution at different instants of time $t$, with $r_0=4,\alpha=1$.}
        \label{fig8}
\end{figure}

The Strong Energy Condition (SEC) is written as,
\begin{equation}
 T_{\mu\nu}u^\mu u^\nu+\frac{1}{2}T\geq 0\implies \frac{\rho+p_r+2p_t}{2}\geq 0,
\end{equation}
which can only be violated if there is an effective repulsive contribution in the energy momentum tensor. For $\alpha=1$, we plot the SEC in Figs. \ref{fig7} and \ref{fig8}.

\begin{figure}[H]
     \centering
    \boxed{ \begin{subfigure}[b]{0.45\textwidth}
         \centering
         \includegraphics[width=\textwidth]{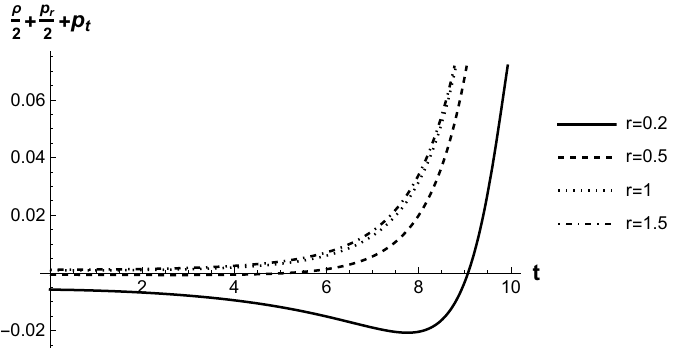}
         \label{fig9a}
     \end{subfigure}
     \hspace{0.2cm}
     \begin{subfigure}[b]{0.45\textwidth}
         \centering
         \includegraphics[width=\textwidth]{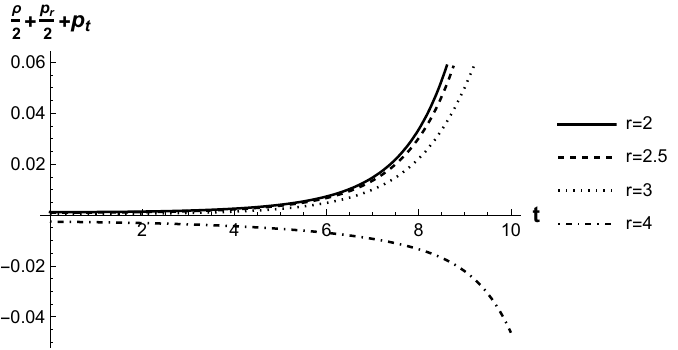}
         \label{fig9b}
     \end{subfigure}}
        \caption{Case I - Study of the SEC during the evolution at different regions characterized by different values of $r$, with $r_0=4,\alpha=2$.}
        \label{fig9}
\end{figure}

\begin{figure}[H]
     \centering
    \boxed{ \begin{subfigure}[b]{0.45\textwidth}
         \centering
         \includegraphics[width=\textwidth]{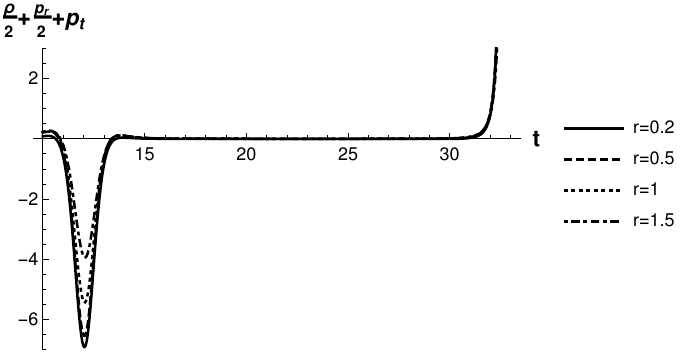}
         \label{fig10a}
     \end{subfigure}
     \hspace{0.2cm}
     \begin{subfigure}[b]{0.45\textwidth}
         \centering
         \includegraphics[width=\textwidth]{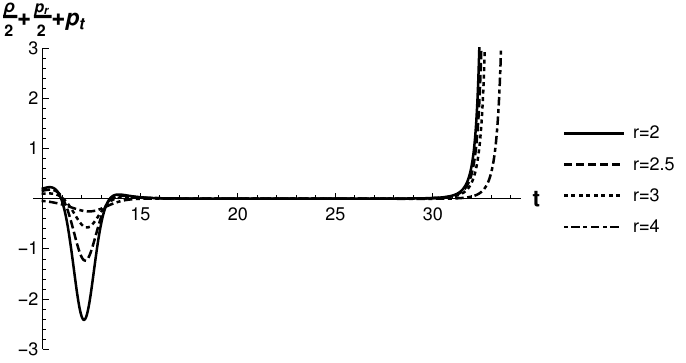}
         \label{fig10b}
     \end{subfigure}}
        \caption{Case I - Study of the SEC during the evolution at different regions characterized by different values of $r$, with $r_0=4,\alpha=2$.}
        \label{fig10}
\end{figure}

\begin{figure}[H]
     \centering
    \boxed{ \begin{subfigure}[b]{0.45\textwidth}
         \centering
         \includegraphics[width=\textwidth]{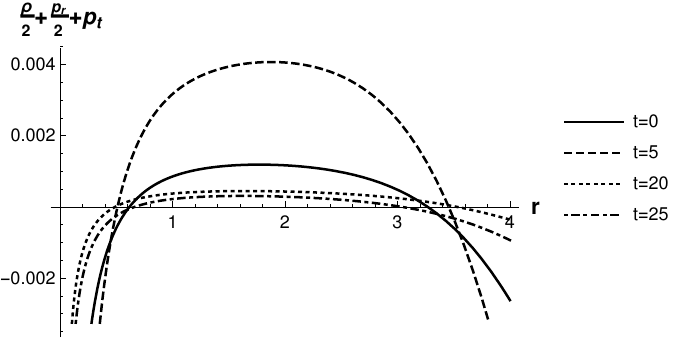}
         \label{fig11a}
     \end{subfigure}
     \hspace{0.2cm}
     \begin{subfigure}[b]{0.45\textwidth}
         \centering
         \includegraphics[width=\textwidth]{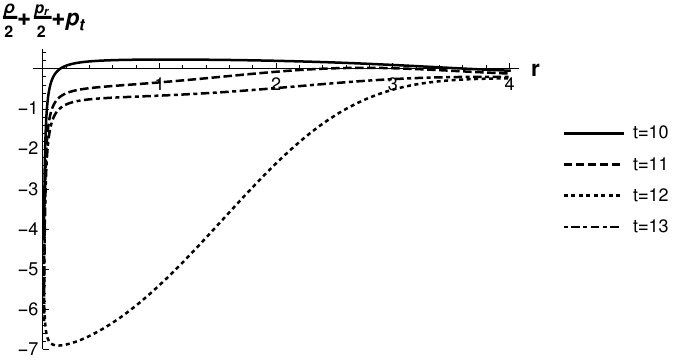}
         \label{fig11b}
     \end{subfigure}}
          \caption{Case I - Study of the SEC within the whole distribution at different instants of time $t$, with $r_0=4,\alpha=2$.}
        \label{fig11}
\end{figure}

A violation of SEC around the central shell as well as close to the boundary is evident. Although it is well satisfied broadly in any other regions inside the collapsing sphere, an implication of repulsive effect due to the quantum corrections is clear. At least for this case, however, this effect is not strong enough to prevent the final formation of singularity. For $\alpha = 2$, we plot the condition in Figs. \ref{fig9}, \ref{fig10}, \ref{fig11}. In Figs. \ref{fig9} and \ref{fig10}, evolution of SEC is shown for two different ranges of time. It can be seen that just like WEC, during a bounce SEC is violated as well. The quantum correction which generates this replulsive effect is not strong enough to stop an eventual re-collapse and a succumbence to singularity.

\begin{figure}[H]
     \centering
    \boxed{ \begin{subfigure}[b]{0.45\textwidth}
         \centering
         \includegraphics[width=\textwidth]{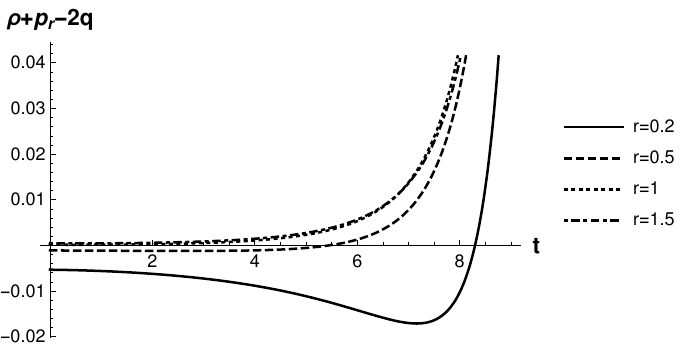}
         \label{fig12a}
     \end{subfigure}
     \hspace{0.2cm}
     \begin{subfigure}[b]{0.45\textwidth}
         \centering
         \includegraphics[width=\textwidth]{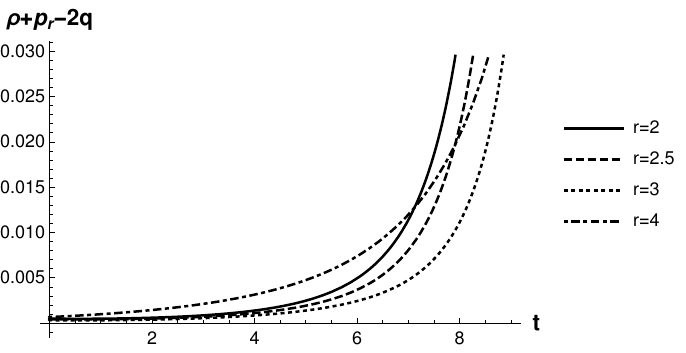}
         \label{fig12b}
     \end{subfigure}}
        \caption{Case I - Study of the NEC during the evolution at different regions characterized by different values of $r$, with $r_0=4,\alpha=1$.}
        \label{fig12}
\end{figure}

\begin{figure}[H]
     \centering
   \boxed{  \begin{subfigure}[b]{0.45\textwidth}
         \centering
         \includegraphics[width=\textwidth]{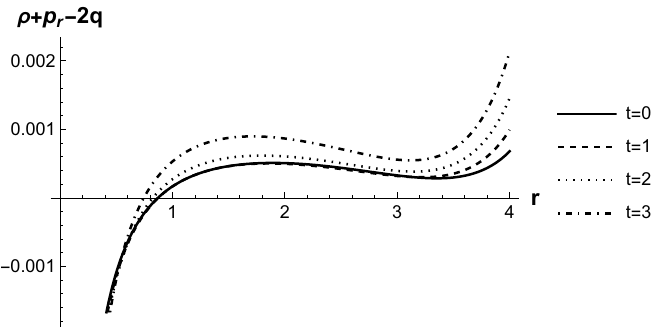}
         \label{fig13a}
     \end{subfigure}
     \hspace{0.2cm}
     \begin{subfigure}[b]{0.45\textwidth}
         \centering
         \includegraphics[width=\textwidth]{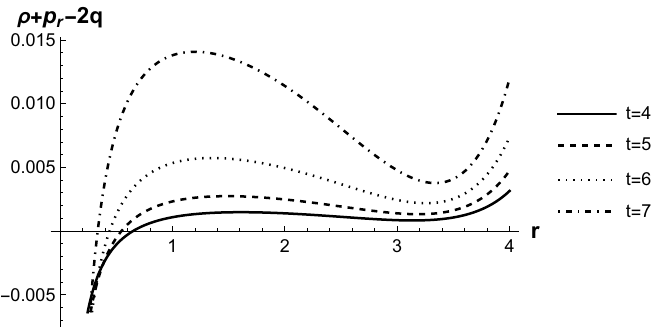}
         \label{fig13b}
     \end{subfigure}}
          \caption{Case I - Study of the NEC within the whole distribution at different instants of time $t$, with $r_0=4,\alpha=1$.}
        \label{fig13}
\end{figure}

For a Null congruence (family of null vectors) characterized by,
\begin{equation}
 k^\mu=\left[\frac{1}{(A+B)f},~ \frac{1}{(A+B)g},~ 0,~ 0\right],
\end{equation}
the NEC is written as
\begin{equation}
 T_{\mu\nu}k^\mu k^\nu\geq 0 \implies ~ \rho+p_\mathrm{r}-2q\geq 0.
\end{equation}
For $\alpha=1$, $NEC \sim \rho+p_\mathrm{r}-2q$ evolutions are shown in Figs. \ref{fig12} and \ref{fig13}.

\begin{figure}[H]
     \centering
  \boxed{   \begin{subfigure}[b]{0.45\textwidth}
         \centering
         \includegraphics[width=\textwidth]{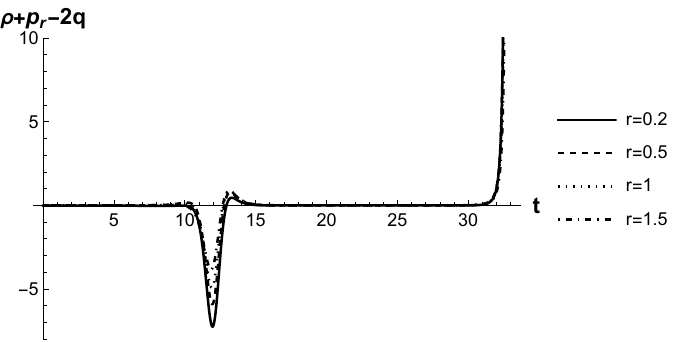}
         \label{fig14a}
     \end{subfigure}
     \hspace{0.2cm}
     \begin{subfigure}[b]{0.45\textwidth}
         \centering
         \includegraphics[width=\textwidth]{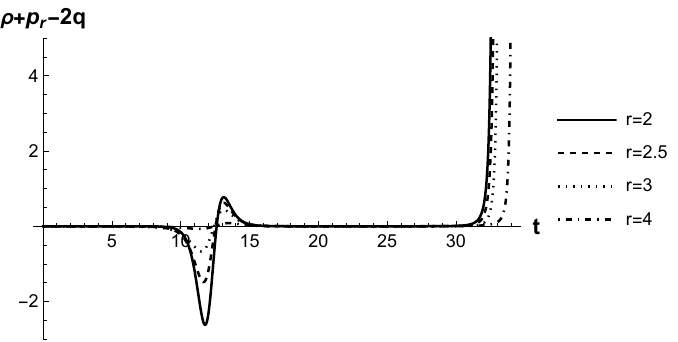}
         \label{fig14b}
     \end{subfigure}}
        \caption{Case I - Study of the NEC during the evolution at different regions characterized by different values of $r$, with $r_0=4,\alpha=2$.}
        \label{fig14}
\end{figure}

\begin{figure}[H]
     \centering
  \boxed{   \begin{subfigure}[b]{0.45\textwidth}
         \centering
         \includegraphics[width=\textwidth]{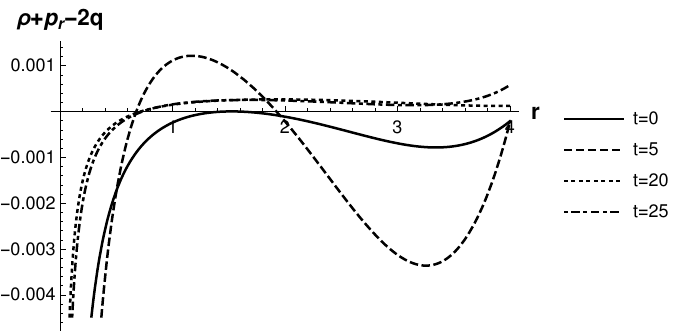}
         \label{fig15a}
     \end{subfigure}
     \hspace{0.2cm}
     \begin{subfigure}[b]{0.45\textwidth}
         \centering
         \includegraphics[width=\textwidth]{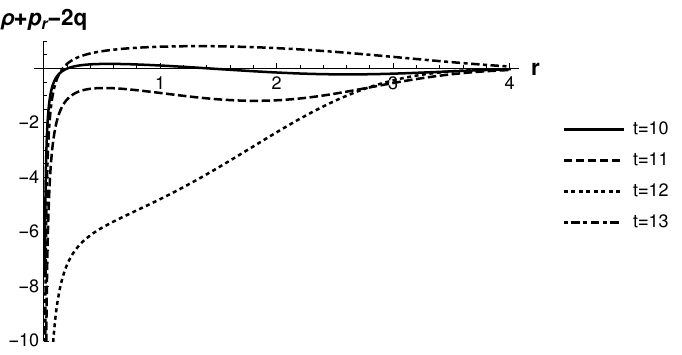}
         \label{fig15b}
     \end{subfigure}}
          \caption{Case I - Study of the NEC within the whole distribution at different instants of time $t$, with $r_0=4,\alpha=2$.}
        \label{fig15}
\end{figure}

These plots suggest that for $\alpha = 1$ the NEC is violated only near central shell of the collpasing sphere. However, for $\alpha = 2$, the evolution is slightly different. This is apparent from Figs. \ref{fig14} and \ref{fig15}, where it is straightforward to note that any bounce is accompanied by the downfall of NEC towards negative values. \\

Qualitatively, the behavior of the collapsing sphere for $\lambda = -1$ is straightforward but non-trivial. The shells obey standard energy conditions broadly, but not always. Any violation of the conditions can be associated with either or both of the following:
quantum corrections playing a dominant role;
a bounce and re-collapse of the shells towards an ultimate singularity.

\subsection{Case II: $\mathbf{\lambda = 1}$}
For a positive $\lambda$ case, we choose the same set of initial conditions:  $T(0) = 10$, $\dot{T}(0) = -0.1$ and the values of the parameters, $\beta = 0.5$ and $\gamma = 0.001$  as in the previous case.
 
\subsubsection{Evolution}
\begin{figure}[H]
     \centering
   \boxed{  \begin{subfigure}[b]{0.45\textwidth}
         \centering
         \includegraphics[width=\textwidth]{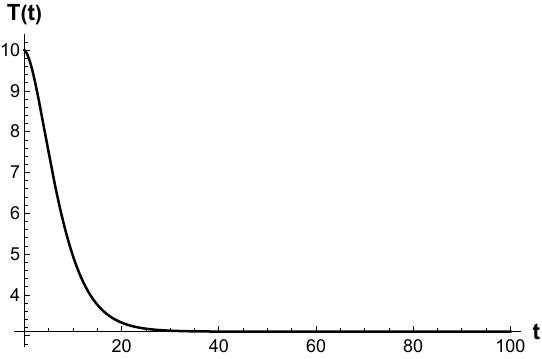}
        \caption{$\alpha=1$}
         \label{fig16a}
     \end{subfigure}
     \hspace{0.2cm}
     \begin{subfigure}[b]{0.45\textwidth}
         \centering
         \includegraphics[width=\textwidth]{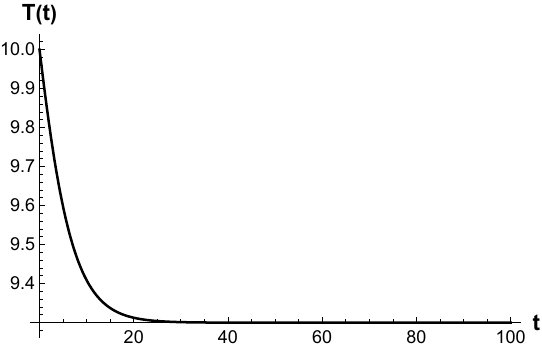}
        \caption{$\alpha=3$}
         \label{fig16b}
     \end{subfigure}}
          \caption{Case II - Evolution of the time-dependent part of the conformal factor, $T$ with $t$ for different values of $\alpha$, where we have chosen $r_0=4$.}
        \label{fig16}
\end{figure}

\begin{figure}[H]
     \centering
   \boxed{  \begin{subfigure}[b]{0.45\textwidth}
         \centering
         \includegraphics[width=\textwidth]{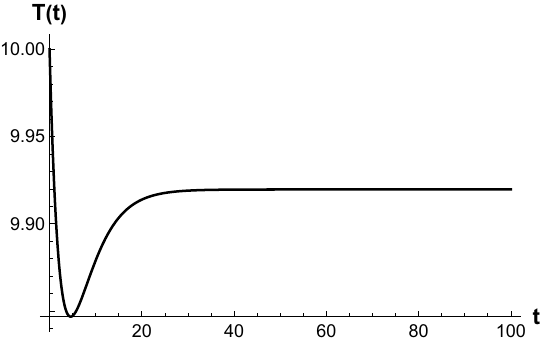}
        \caption{$\alpha=3.2$}
         \label{fig17a}
     \end{subfigure}
     \hspace{0.2cm}
     \begin{subfigure}[b]{0.45\textwidth}
         \centering
         \includegraphics[width=\textwidth]{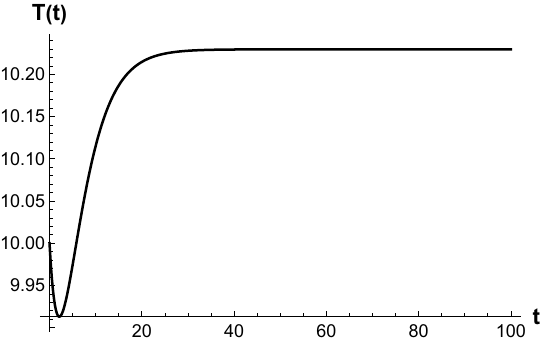}
         \caption{$\alpha=3.3$}
         \label{fig17b}
     \end{subfigure}}
          \caption{Case II - Evolution of the time-dependent part of the conformal factor, $T$ with $t$ for different values of $\alpha$, where we have chosen $r_0=4$.}
        \label{fig17}
\end{figure}

The evolution of $T(t)$ for $r_0 = 4$ is shown in Fig. \ref{fig16} and \ref{fig17}. Initially, the proper volume of the sphere shrinks with pace. Howver, the main qualitative difference from a positive $\lambda$ case is that it never reaches a zero proper volume at a finite time. Asymptotically the geometry is represented by a constant (non-zero) conformal factor over the quantum corrected Scwarzschild interior. The final value of $T(t)$ depends on the choice of $\alpha$. There also exists a specific value of $\alpha = \alpha_0$ such that whenever $\alpha > \alpha_0$ a collapse-and-bounce scenario is observed. The functon $T(t)$ keeps increasing after the bounce until it simply gives a constant value for the conformal factor. The value of $\alpha$ also determines whether the final volume shall be greater or lesser than the initial volume. This can be understood from Figs. \ref{fig17a} dnd \ref{fig17b}.

\begin{figure}[H]
     \centering
 \boxed{    \begin{subfigure}[b]{0.45\textwidth}
         \centering
         \includegraphics[width=\textwidth]{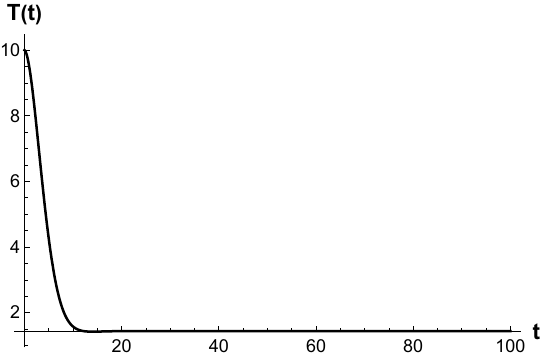}
         \label{fig18a}
     \end{subfigure}
     \hspace{0.2cm}
     \begin{subfigure}[b]{0.45\textwidth}
         \centering
         \includegraphics[width=\textwidth]{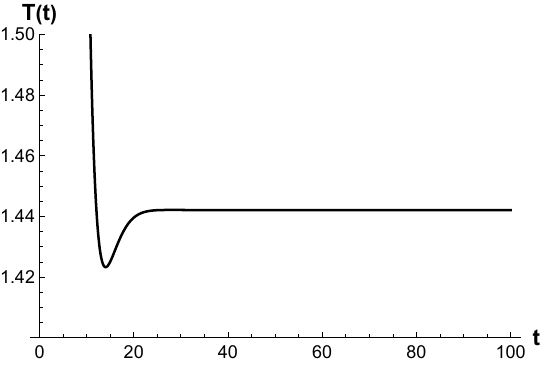}
         \label{fig18b}
     \end{subfigure}}
     \caption{Case II - Evolution of the time-dependent part of the conformal factor, $T$ with $t$ for $\alpha=1$, and with $r_0=2$.}
        \label{fig18}
\end{figure}

Unlike the first case, here, the qualitative nature of the evolution depends on the initial size of the sphere, i.e., value of $r_0$. We plot the evolution for $r_0 = 2$ and $\alpha = 1$ in Fig. \ref{fig18}. For a sphere whose boundary is at $r_0 = 2$, irrespective of the value of $\alpha$ we choose, there is always a bounce after which the proper volume approaches a constant value.

We infer that for $\lambda=1$, the evolution always leads to non-singular end states. To have insights about different collapsing and bouncing shells we study the validity of different energy conditions for different shells, as a function of time.

\subsubsection{Energy conditions}
\begin{figure}[H]
     \centering
   \boxed{  \begin{subfigure}[b]{0.45\textwidth}
         \centering
         \includegraphics[width=\textwidth]{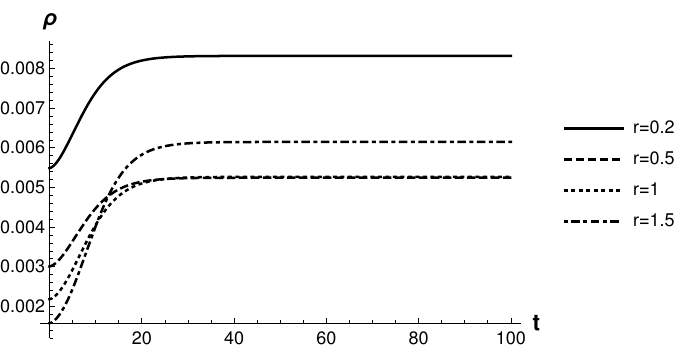}
         \label{fig19a}
     \end{subfigure}
     \hspace{0.2cm}
     \begin{subfigure}[b]{0.45\textwidth}
         \centering
         \includegraphics[width=\textwidth]{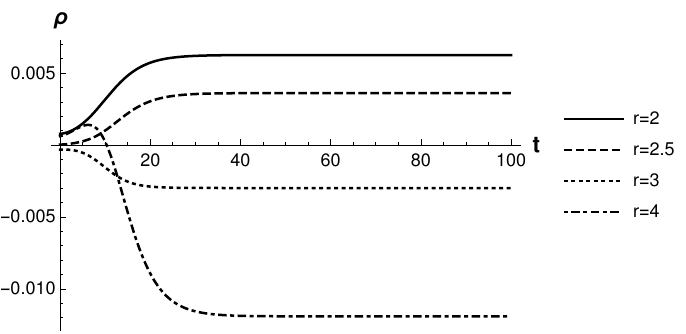}
         \label{fig19b}
     \end{subfigure}}
          \caption{Case II - Evolution of the energy densities for different regions within the distribution with time for $\alpha=1$ and $r_0=4$.}
        \label{fig19}
\end{figure}

We consider the WEC for $\alpha = 1$ and plot it as a function of time in Fig. \ref{fig19}. The plot indicates that after a certain time the energy density of the distribution approaches a non-zero constant value. This is indeed expected from the nature of the evolution.

\begin{figure}[H]
     \centering
   \boxed{  \begin{subfigure}[b]{0.45\textwidth}
         \centering
         \includegraphics[width=\textwidth]{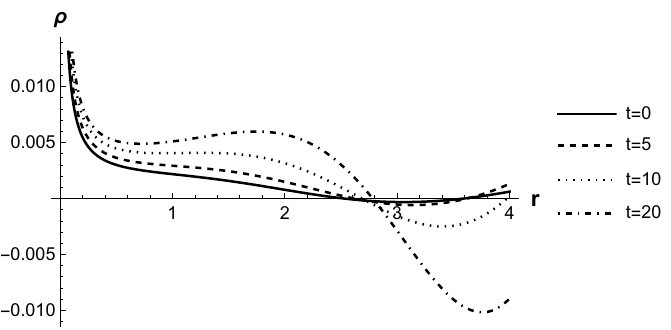}
         \label{fig20a}
     \end{subfigure}
     \hspace{0.2cm}
     \begin{subfigure}[b]{0.45\textwidth}
         \centering
         \includegraphics[width=\textwidth]{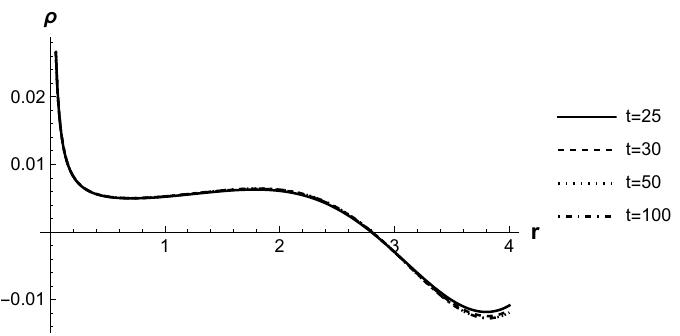}
         \label{fig20b}
     \end{subfigure}}
          \caption{Case II - Radial dependence of the energy densities at different instances of time for $\alpha=1$ and $r_0=4$.}
        \label{fig20}
\end{figure}

The radial dependence of the energy density profiles are shown in Fig. \ref{fig20}, for different time snapshots, i.e., different values of time. The energy densities of the regions near the boundary hypersurface remains negative most of the time. We can attribute this nature to the quantum corrections in the metric tensor. Far from the boundary, energy densities are always positive.

\begin{figure}[H]
     \centering
   \boxed{  \begin{subfigure}[b]{0.45\textwidth}
         \centering
         \includegraphics[width=\textwidth]{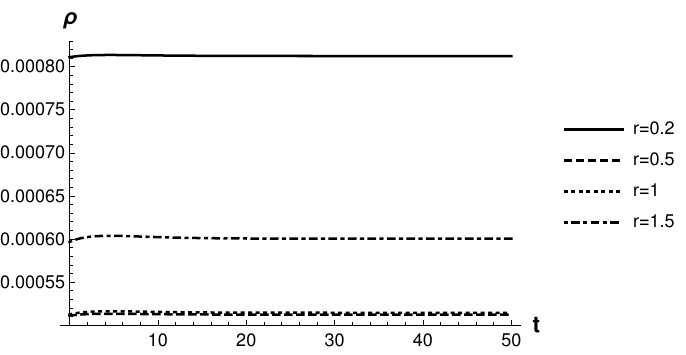}
         \label{fig21a}
     \end{subfigure}
     \hspace{0.2cm}
     \begin{subfigure}[b]{0.45\textwidth}
         \centering
         \includegraphics[width=\textwidth]{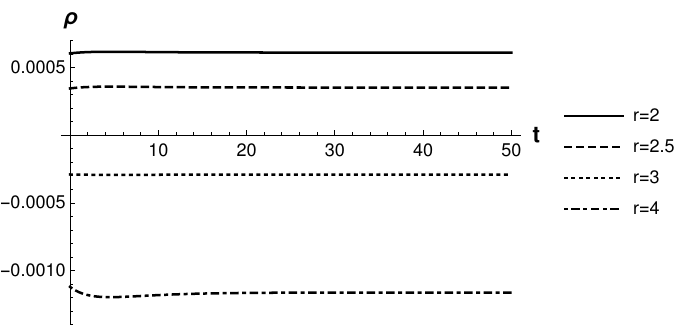}
         \label{fig21b}
     \end{subfigure}}
          \caption{Case II - Evolution of the energy densities for different regions within the distribution with time in the case of $\alpha=3.2$ and $r_0=4$.}
        \label{fig21}
\end{figure}

\begin{figure}[H]
     \centering
     \boxed{\begin{subfigure}[b]{0.45\textwidth}
         \centering
         \includegraphics[width=\textwidth]{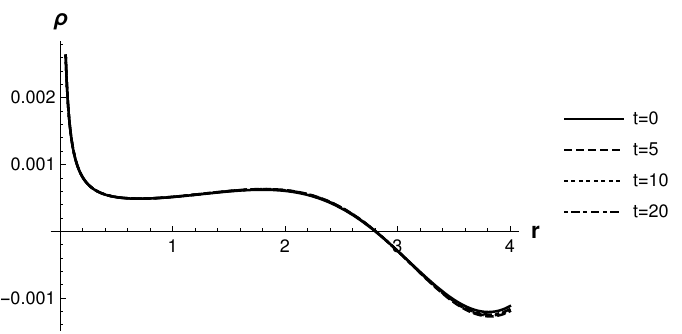}
        \caption{$r=0.2$}
         \label{fig22a}
     \end{subfigure}
     \hspace{0.2cm}
     \begin{subfigure}[b]{0.45\textwidth}
         \centering
         \includegraphics[width=\textwidth]{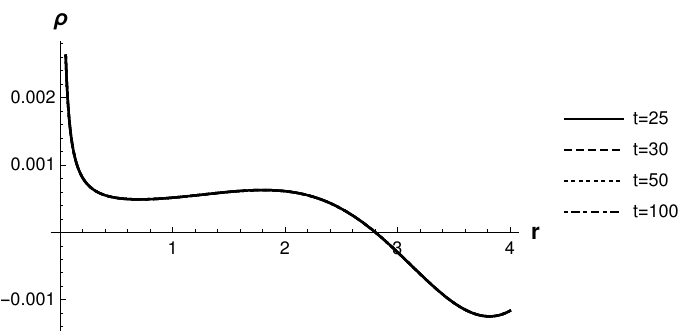}
         \caption{$r=4$}
         \label{fig22b}
     \end{subfigure}}
          \caption{Case II - Radial dependence of the energy densities at different instances of time for $\alpha=3.2$ and $r_0=4$.}
        \label{fig22}
\end{figure}

We also demonstrate the situation for $\alpha=3.2$ which can be considered as a proto-type of evolutions with a bounce. The WEC evolution is shown in the Figs. \ref{fig21} and \ref{fig22}. The qualitative inference is that the energy densities of the interior regions near the centre of the collapsing core slowly increases with the evolution and then approaches a constant value. On the other hand, energy densities of regions near the boundary decreases before appproaching the constant value. We also plot the energy density profiles for $\alpha = 3.2$ at different instances of time in Fig. \ref{fig22}. As expected we find that that WEC is always obeyed in the regions near the center, whereas it is violated in regions near the boundary.


\begin{figure}[H]
     \centering
    \boxed{ \begin{subfigure}[b]{0.45\textwidth}
         \centering
         \includegraphics[width=\textwidth]{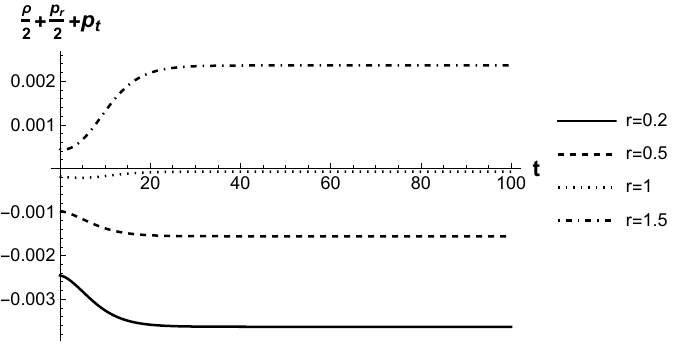}
         \label{fig23a}
     \end{subfigure}
     \hspace{0.2cm}
     \begin{subfigure}[b]{0.45\textwidth}
         \centering
         \includegraphics[width=\textwidth]{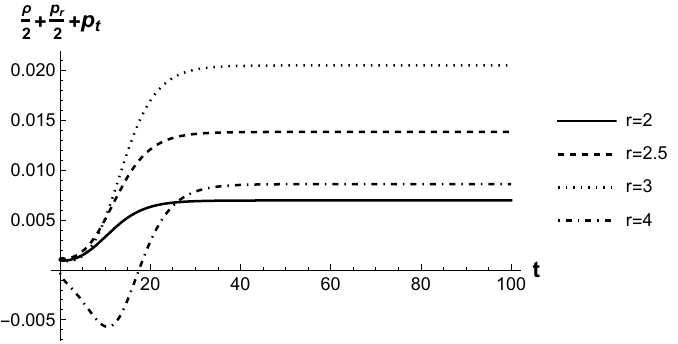}
         \label{fig23b}
     \end{subfigure}}
          \caption{Case II - Evolution of the quantity, $\frac{\rho+p_\mathrm{r}+2p_\mathrm{t}}{2}$ for different regions within the distribution in the case of $\alpha=1$ and $r_0=4$}
        \label{fig23}
\end{figure}

\begin{figure}[H]
     \centering
  \boxed{   \begin{subfigure}[b]{0.45\textwidth}
         \centering
         \includegraphics[width=\textwidth]{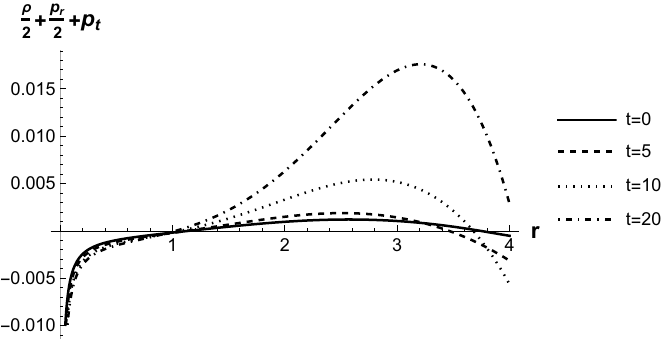}
         \label{fig24a}
     \end{subfigure}
     \hspace{0.2cm}
     \begin{subfigure}[b]{0.45\textwidth}
         \centering
         \includegraphics[width=\textwidth]{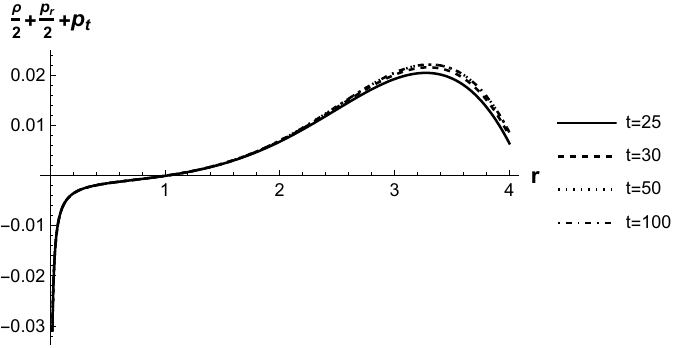}
         \label{fig24b}
     \end{subfigure}}
          \caption{Case II - Radial dependence of the quantity, $\frac{\rho+p_\mathrm{r}+2p_\mathrm{t}}{2}$ at different instances of time in the case of $\alpha=1$ and $r_0=4$.}
        \label{fig24}
\end{figure}

We proceed and investigate the nature of SEC for this spherical system with $\alpha > 0$. For $\alpha = 1$, the evolution of $\frac{\rho + p_\mathrm{r} + 2 p_\mathrm{t}}{2}$ is given in Figs. \ref{fig23} and \ref{fig24}. It clearly suggests that SEC is violated during the collapse, near the collapsing core ($r \rightarrow 0$). This results in an effective negative pressure contribution (quantum effects inspired) which prevents a formation of singularity.

\begin{figure}[H]
     \centering
  \boxed{   \begin{subfigure}[b]{0.45\textwidth}
         \centering
         \includegraphics[width=\textwidth]{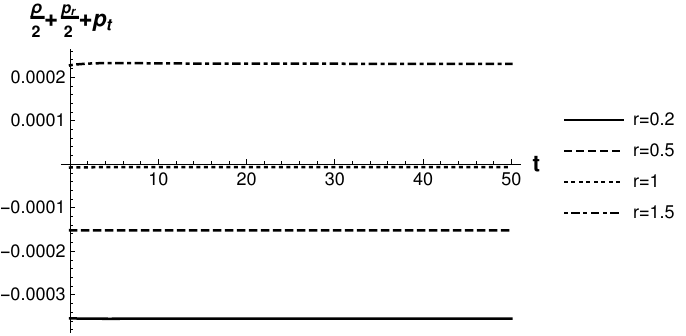}
         \label{fig25a}
     \end{subfigure}
     \hspace{0.2cm}
     \begin{subfigure}[b]{0.45\textwidth}
         \centering
         \includegraphics[width=\textwidth]{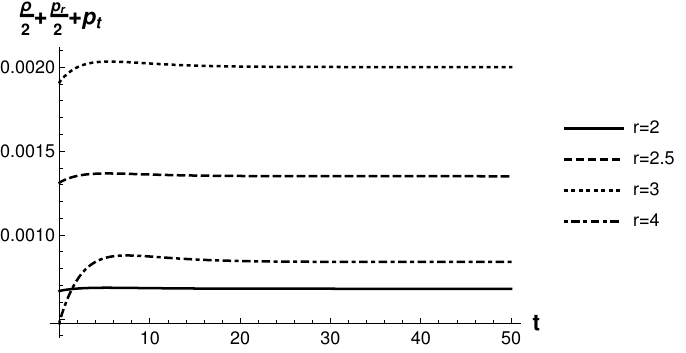}
         \label{fig25b}
     \end{subfigure}}
          \caption{Case II - Evolution of the quantity, $\frac{\rho+p_\mathrm{r}+2p_\mathrm{t}}{2}$ for different regions within the distribution in the case of $\alpha=3.2$ and $r_0=4$.}
        \label{fig25}
\end{figure}

\begin{figure}[H]
     \centering
  \boxed{   \begin{subfigure}[b]{0.45\textwidth}
         \centering
         \includegraphics[width=\textwidth]{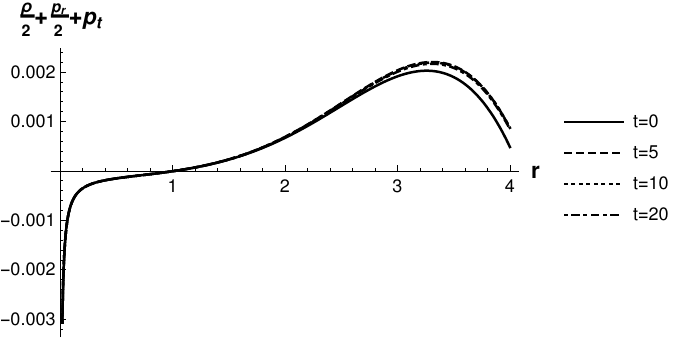}
         \label{fig25a}
     \end{subfigure}
     \hspace{0.2cm}
     \begin{subfigure}[b]{0.45\textwidth}
         \centering
         \includegraphics[width=\textwidth]{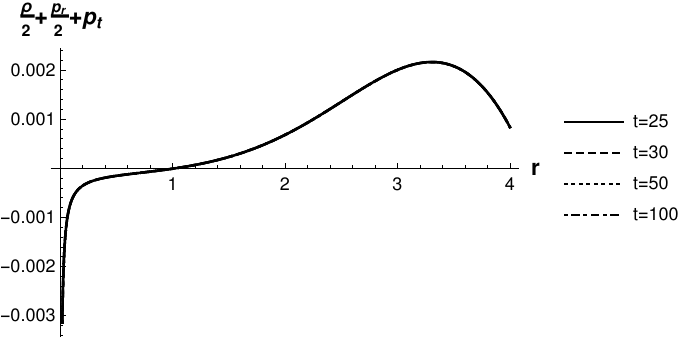}
         \label{fig25b}
     \end{subfigure}}
          \caption{Case II - Radial dependence of the quantity, $\frac{\rho+p_\mathrm{r}+2p_\mathrm{t}}{2}$ at different instances of time in the case of $\alpha=3.2$ and $r_0=4$.}
        \label{fig26}
\end{figure}

For $\alpha = 3.2$, we have a non-singular bouncing solution, and the evolution of SEC (see Figs. \ref{fig25} and \ref{fig26}) shows clear violation near the center of the collapsing sphere.

\begin{figure}[H]
     \centering
 \boxed{    \begin{subfigure}[b]{0.45\textwidth}
         \centering
         \includegraphics[width=\textwidth]{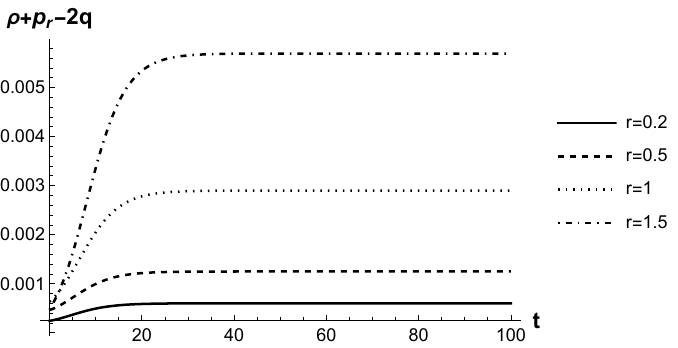}
         \label{fig27a}
     \end{subfigure}
     \hspace{0.2cm}
     \begin{subfigure}[b]{0.45\textwidth}
         \centering
         \includegraphics[width=\textwidth]{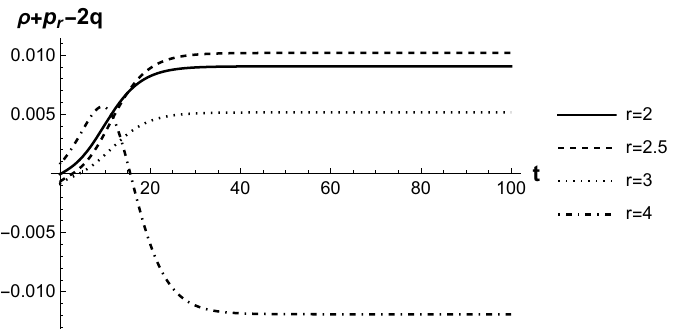}
         \label{fig27b}
     \end{subfigure}}
          \caption{Case II - Evolution of the quantity, $\rho+p_\mathrm{r}-2q$ for different regions within the distribution in the case of $\alpha=3.2$ and $r_0=4$.}
        \label{fig27}
\end{figure}

\begin{figure}[H]
     \centering
   \boxed{  \begin{subfigure}[b]{0.45\textwidth}
         \centering
         \includegraphics[width=\textwidth]{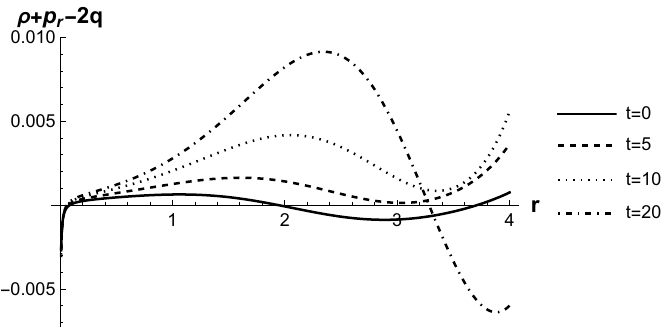}
         \label{fig28a}
     \end{subfigure}
     \hspace{0.2cm}
     \begin{subfigure}[b]{0.45\textwidth}
         \centering
         \includegraphics[width=\textwidth]{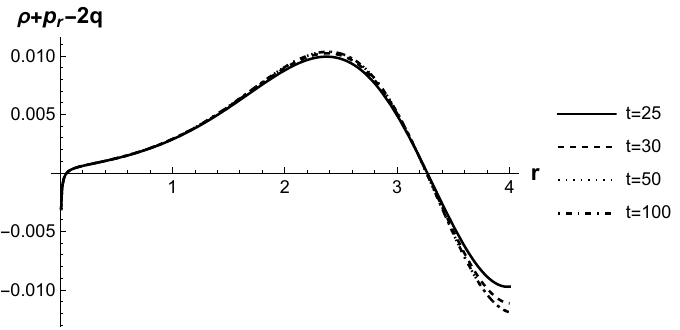}
         \label{fig28b}
     \end{subfigure}}
          \caption{Case II - Radial dependence of the quantity, $\rho+p_\mathrm{r}-2q$ at different instances of time in the case of $\alpha=1$ and $r_0=4$.}
        \label{fig28}
\end{figure}

\begin{figure}[H]
     \centering
  \boxed{   \begin{subfigure}[b]{0.45\textwidth}
         \centering
         \includegraphics[width=\textwidth]{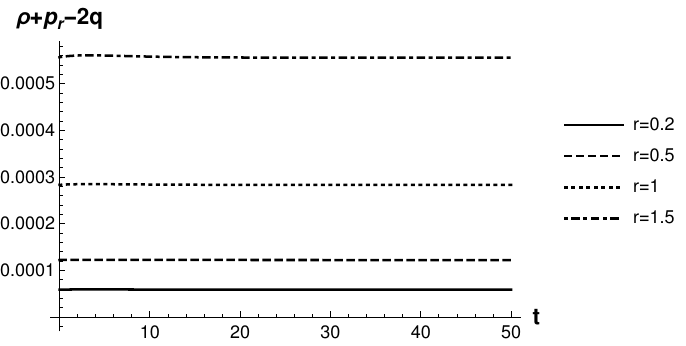}
         \label{fig29a}
     \end{subfigure}
     \hspace{0.2cm}
     \begin{subfigure}[b]{0.45\textwidth}
         \centering
         \includegraphics[width=\textwidth]{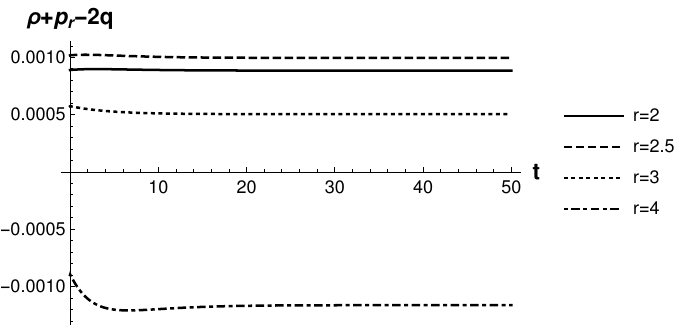}
         \label{fig29b}
     \end{subfigure}}
         \caption{Case II - Evolution of the quantity, $\rho+p_\mathrm{r}-2q$ for different regions within the distribution in the case of $\alpha=3.2$ and $r_0=4$.}
        \label{fig29}
\end{figure}

\begin{figure}[H]
     \centering
   \boxed{  \begin{subfigure}[b]{0.45\textwidth}
         \centering
         \includegraphics[width=\textwidth]{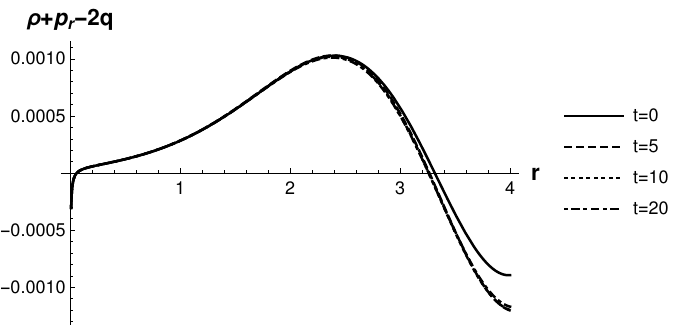}
       \caption{$r=0.2$}
         \label{fig30a}
     \end{subfigure}
     \hspace{0.2cm}
     \begin{subfigure}[b]{0.45\textwidth}
         \centering
         \includegraphics[width=\textwidth]{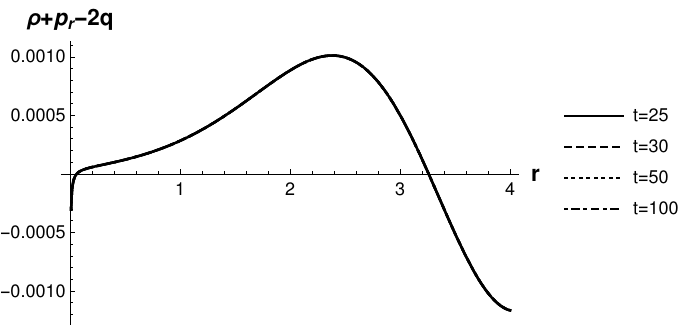}
         \caption{$r=4$}
         \label{fig30b}
     \end{subfigure}}
         \caption{Case II - Radial dependence of the quantity, $\rho+p_\mathrm{r}-2q$ at different instances of time in the case of $\alpha=3.2$ and $r_0=4$.}
        \label{fig30}
\end{figure}

Finally, we plot the evolution of $\rho+p_\mathrm{r}-2q$ (NEC) for $\alpha = 1$ in Figs. \ref{fig27} and \ref{fig28}. We can see a clear violation of the NEC in regions very close to the center. Also, the NEC is always violated in the regions near the boundary at later stages of the evolution. For $\alpha = 3.2$, it is easy to infer something similar refering to the Figs. \ref{fig29} and \ref{fig30}.

In a nutshell, the behavior of the collapsing sphere for $\lambda = 1$ is quite interesting. The collapsing shells violate the SEC and NEC near the collapsing core but obey the WEC there. A violation of these conditions can be associated with either or both of the following:
quantum corrections playing a dominant role;
a bounce of the clustered matter due to the generation of an effective negative pressure contribution.

One can claim that the radial dependence of the conformal factor, i.e., $B(r)$, may be thought of as having a quantum-inspired genesis. We finish this section by providing an additional evidence to this claim. The fact is, for $B(r)=0$, the $T(t)$ evolution equation can be written as

\begin{equation}\label{Teq1}
\frac{2\ddot{T}}{T}-\left(\frac{\dot{T}}{T}\right)^2-\frac{2x_2\dot{T}}{T}-x_5=0,
\end{equation}
which has the solution,

\begin{equation}
T(t)= {C_1} e^{{x_2}t} \cos ^2\left(\frac{1}{2} t \sqrt{-{x_2}^2-{x_5}}-{C_2} \sqrt{-{x_2}^2-{x_5}}\right).
\end{equation}

Without any loss of generality we can choose $C_{1} = 1$ and $C_{2} = 0$. Therefore the conformal factor for this special case can be written as
\begin{equation}
 A(t) = e^{{x_2}t} \cos ^2\left(\frac{1}{2} t \sqrt{-{x_2}^2-{x_5}}\right),
\end{equation}
which is real when ${x_2}^2+{x_5}\leq 0$. Now, at $t \rightarrow \frac{\pi}{\sqrt{-{x_2}^2-{x_5}}}$, $A(t)\rightarrow 0$ and therefore the evolution hits a singularity. The possibility of a bounce or any other non-singular fate is removed. Qualitatively, we will not be able to construct a collapsing sphere ending up in a geometry conformal to the quantum-corrected black hole solution as in Eq. \eqref{qcorsh}. If and only if $B(r)\neq 0$, a bounce can occur under suitable initial conditions. The function $B(r)$ leads to a special kind of inhomogeneity in the metric coefficients, the four velocity of the comoving observer and naturally, to the energy-momentum tensor components. As can be seen, this contribution is crucial to manifest repulsive quantum effects during the collapse.

\section{Formation of a Black Hole Singularity for $\lambda < 0$}
Since the negative $\lambda$ case always produces a singularity, irrespective of the choices of $\alpha$ and $r_0$, we check the visibility of this singular end-state. On a technical note, a \textit{visibility} means the scope to allow an exchange of matter or radiation with a faraway observer. This can be decided by investigating if the null surface formation condition is satisfied anywhere throughout the collapse. The condition is written as,
\begin{equation}
g^{\mu\nu}Y_{,\mu}Y_{,\nu} = 0,
\end{equation}
where $Y(r,t)$ is the radius of the two-sphere, given by $r\lbrace A(t) + B(r) \rbrace$ in our construction. For the metric as in Eq. (\ref{qcorsh}) we write the above equation as,
\begin{equation}
\left(\frac{r\dot{A}}{f}\right)^2 -  \left(\frac{ A + B + rB'}{g}\right)^2=0.
\end{equation}
To have a picture regarding the formation of an apparent horizon, we examine if the quantity in the left hand side becomes zero at any stage during the collapse. An important thing to note at this stage is that the central shell becomes singular at first and subsequently the other shells. Let us see if the apparent horizon forms before the central shell hits the singularity. 

\begin{figure}[H]
  \centering
  \boxed{ \begin{subfigure}[b]{0.45\textwidth}\centering \includegraphics[width=\textwidth]{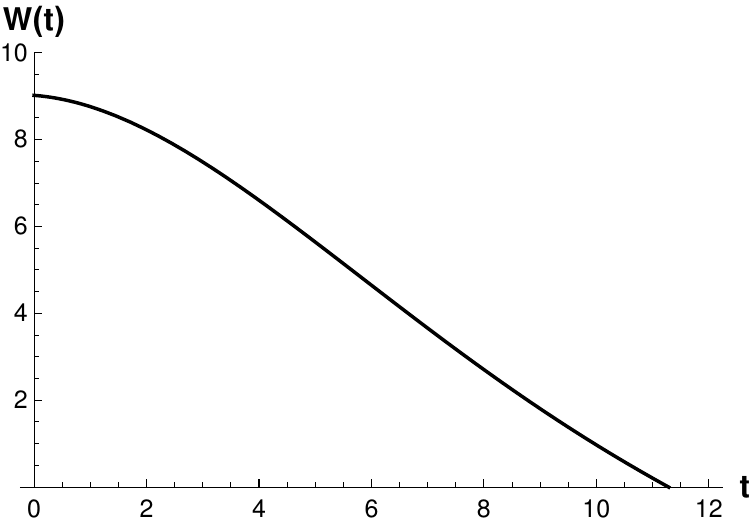}
           \caption{$\alpha=1$}
           \label{figcenta}
          \end{subfigure}\hspace{0.2cm}
\begin{subfigure}[b]{0.45\textwidth}
         \centering
         \includegraphics[width=\textwidth]{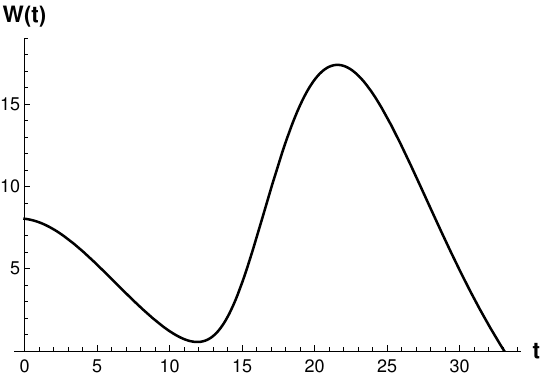}
       \caption{$\alpha=2$}
         \label{figcentb}
     \end{subfigure}}
       \caption{Plot of $W(t)$ for the central shell for the case $\lambda=-1$ with different values of $\alpha$}\label{figcent}
         \end{figure}

\begin{figure}[H]
     \centering
  \fbox{\parbox{0.95\textwidth}{ \begin{subfigure}[b]{0.3\textwidth}
         \centering
         \includegraphics[width=\textwidth]{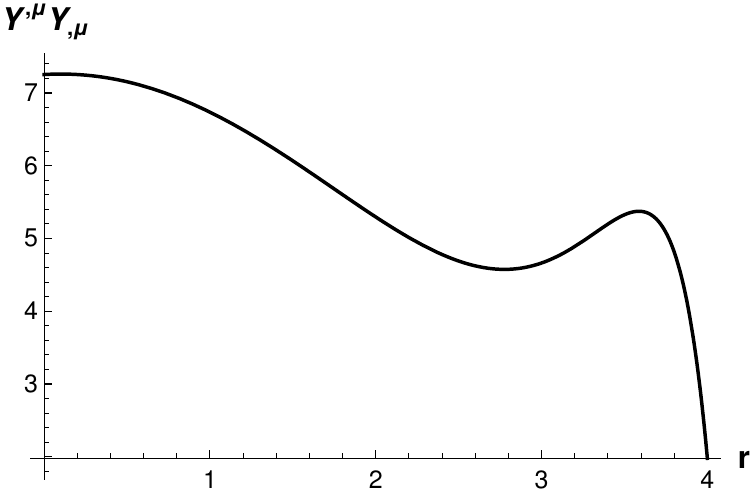}
       \caption{$t=8$}
         \label{figAHa}
     \end{subfigure}
     \hspace{0.2cm}
     \begin{subfigure}[b]{0.3\textwidth}
         \centering
         \includegraphics[width=\textwidth]{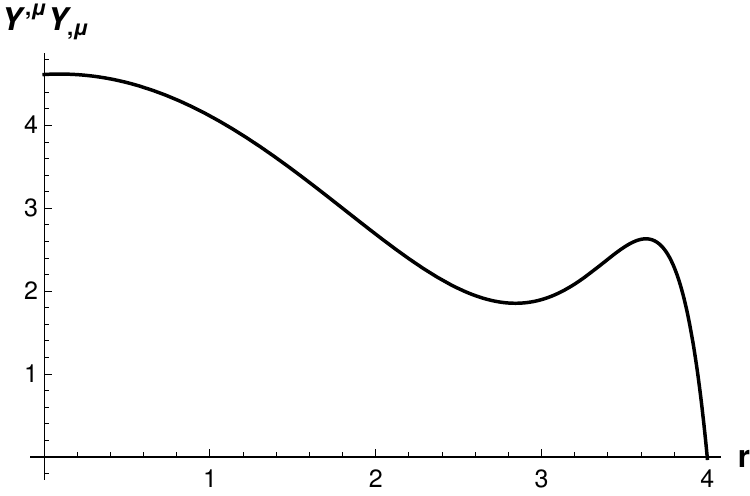}
         \caption{$t=8.6$}
         \label{figAHb}
     \end{subfigure}
     \hspace{0.2cm}
     \begin{subfigure}[b]{0.3\textwidth}
         \centering
         \includegraphics[width=\textwidth]{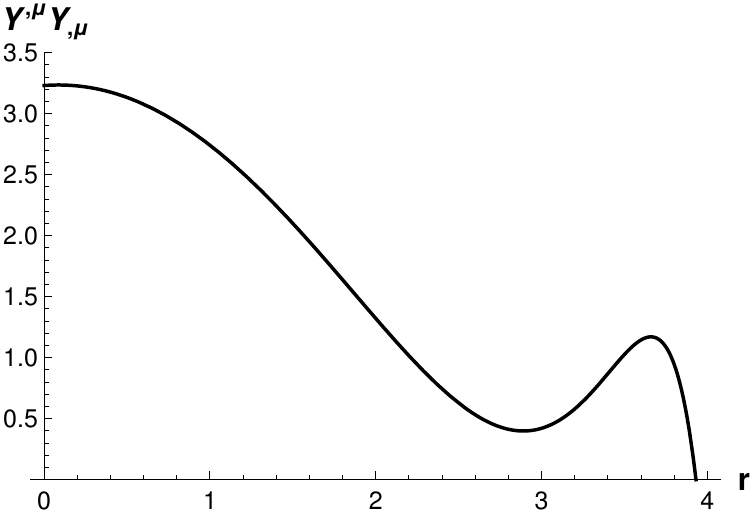}
         \caption{$t=9$}
         \label{figAHc}
     \end{subfigure}
     
     \begin{subfigure}[b]{0.3\textwidth}
         \centering
         \includegraphics[width=\textwidth]{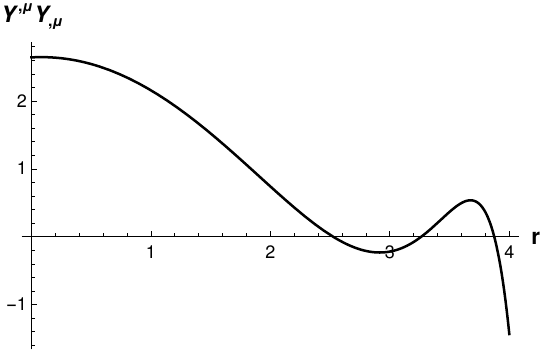}
       \caption{$t=9.2$}
         \label{figAHd}
     \end{subfigure}
     \hspace{0.2cm}
     \begin{subfigure}[b]{0.3\textwidth}
         \centering
         \includegraphics[width=\textwidth]{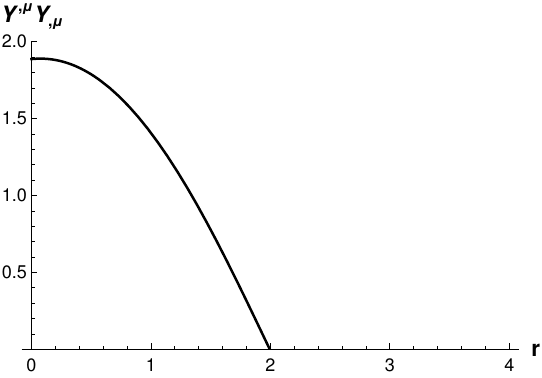}
       \caption{$t=9.5$}
         \label{figAHe}
     \end{subfigure}\hspace{0.2cm}
     \begin{subfigure}[b]{0.3\textwidth}
         \centering
         \includegraphics[width=\textwidth]{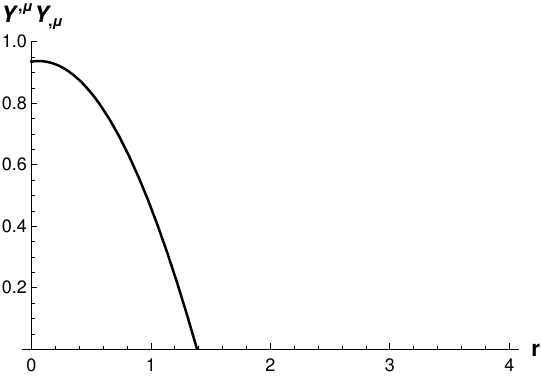}
       \caption{$t=10$}
         \label{figAHf}
     \end{subfigure}}}
         \caption{Plot of $g^{\mu\nu}Y_{,\mu}Y_{,\nu}$ with $r$ at different values of $t$ for $\lambda=-1$ and $\alpha=1$.}
        \label{figAH}
\end{figure}

The evolution of the conformal factor $W(t)\equiv A(t)+B(0)$ (i.e. at $r=0$) for $\alpha=1$  is shown in Fig. \ref{figcenta}. The central shell hits the singularity near $t=11$. After a careful study, we have found that in this case the quantity $g^{\mu\nu}Y_{,\mu}Y_{,\nu}$ becomes zero for a value of $r\in [0,4]$ after a certain time. This signifies formation of the apparent horizon. To illustrate it more clearly we plot the quantity $g^{\mu\nu}Y_{,\mu}Y_{,\nu}$ with $r$ at $t=8$, $8.6$, $9$, $9.2$, $9.5$ and $10$  in Fig. \ref{figAH}. The plots suggest that the apparent horizon forms very close to the boundary of the distribution at $t=8.6$ (there is no apparent horizon formation until $t=8.5$) after which this null sphere shrinks. Then three horizons appear simultaneously during the evolution and finally the system contain only one apparent horizon. Hence the singularity is always enveloped by a horizon and is not visible. This kind of triple horizon has been found to appear in cosmological black holes under the framework of a certain class of modified theories \cite{Saghafi2023}.

\begin{figure}[H]
     \centering
  \fbox{ \parbox{0.95\textwidth} {\begin{subfigure}[b]{0.3\textwidth}
         \centering
         \includegraphics[width=\textwidth]{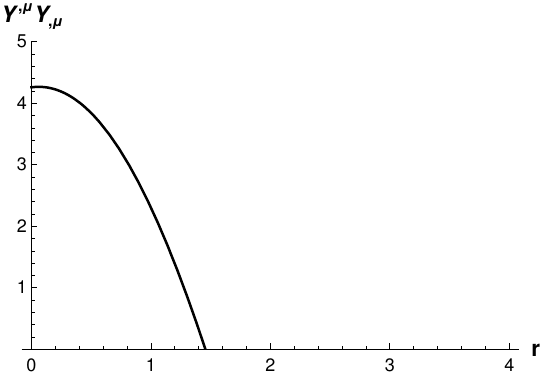}
       \caption{$t=14$}
         \label{figAHbouncea}
     \end{subfigure}
     \hspace{0.2cm}
     \begin{subfigure}[b]{0.3\textwidth}
         \centering
         \includegraphics[width=\textwidth]{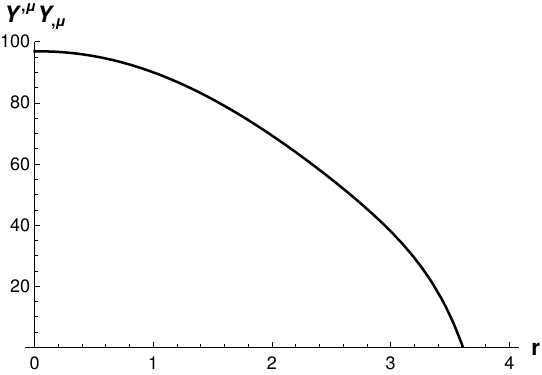}
         \caption{$t=17$}
         \label{figAHbounceb}
     \end{subfigure}
     \hspace{0.2cm}
     \begin{subfigure}[b]{0.3\textwidth}
         \centering
         \includegraphics[width=\textwidth]{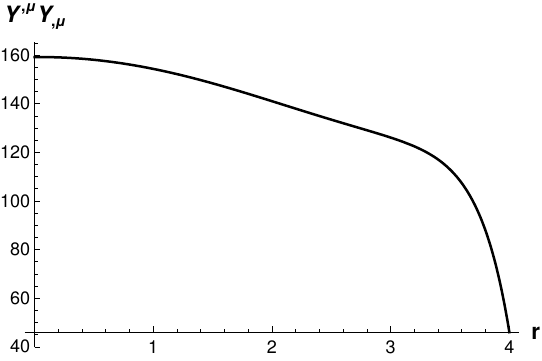}
         \caption{$t=18$}
         \label{figAHbouncec}
     \end{subfigure}
     
    \begin{subfigure}
      {0.3\textwidth}
         \centering
         \includegraphics[width=\textwidth]{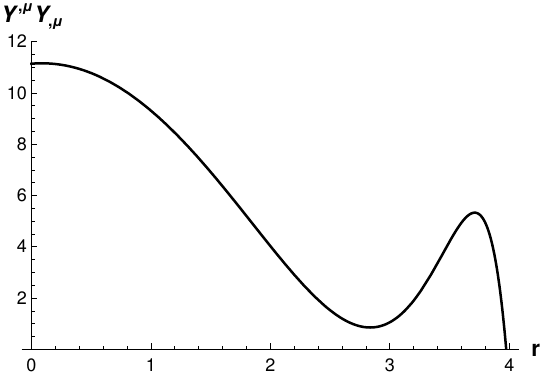}
         \caption{$t=30.9$}
         \label{figAHbounced}
     \end{subfigure}
     \hspace{0.2cm}
     \begin{subfigure}
      {0.3\textwidth}
         \centering
         \includegraphics[width=\textwidth]{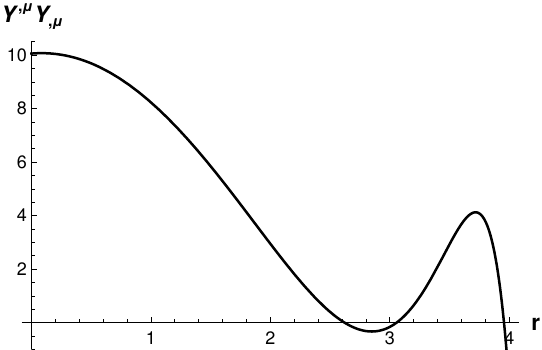}
         \caption{$t=31$}
         \label{figAHbouncee}
     \end{subfigure}
     \hspace{0.2cm}
     \begin{subfigure}
      {0.3\textwidth}
         \centering
         \includegraphics[width=\textwidth]{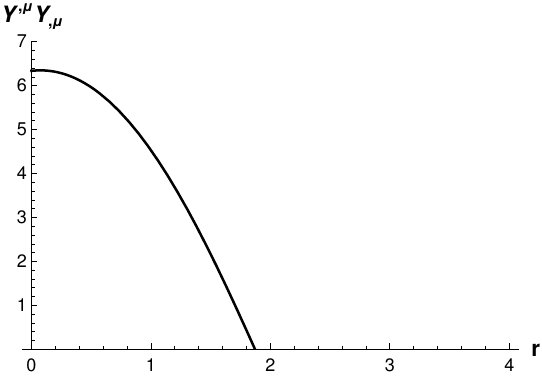}
         \caption{$t=31.4$}
         \label{figAHbouncef}
     \end{subfigure}}}
         \caption{Plot of $g^{\mu\nu}Y_{,\mu}Y_{,\nu}$ with $r$ at different values of $t$ for $\lambda=-1$ and $\alpha=1$.}
        \label{AHalpha2}
\end{figure}

For $\alpha=2$ the situation is more intriguing. The evolution of the quantity  $g^{\mu\nu}Y_{,\mu}Y_{,\nu}$ in this scenario is depicted through the sequence of plots in Fig. \ref{AHalpha2}. An apparent horizon appears after the bounce (Fig. \ref{figAHbouncea}), increases in size (Fig. \ref{figAHbounceb}) and disappears (Fig. \ref{figAHbouncec}). Another horizon again appears during the second stage of contraction (Fig. \ref{figAHbounced}). After that, three exist three dynamical horizons (Fig. \ref{figAHbouncee}) at a certain stage of the collapse. Two of them disappear at a later time and finally the system contains a single horizon (Fig. 
\ref{figAHbouncef}) enveloping the singularity. It is worthwhile to mention that formation and disappearance of dynamical horizons through generation of shock waves for several interesting systems incorporating quantum effects have been discussed quite extensively in a few recent works \cite{PhysRevLett.128.121301, PhysRevD.106.024014, PhysRevD.106.046012}.

\section{Formation of a Wormhole Throat for $\lambda > 0$}
A violation of energy conditions is not unheard of in modified theories of gravity, in particular, for non-minimally coupled theories \cite{BARCELO1999127, PhysRevD.54.6233}. In particular the entire precinct of Dark Energy being able to drive an accelerated expansion of the universe lies on the notion of a violated strong energy condition. However, the null energy condition in particular drives our curiosity to check if the metric as in Eq. (\ref{intmet}), behaves similar to a wormhole geometry during the course of the evolution, i.e., for any arbitrary value of $t$. The most prominent feature of a wormhole is a throat, which connects distant parts of a geometry (or even two different metrics resembling two different universe!) through a tunnel. We write the necessary and sufficient conditions for a wormhole geometry (can also be called a \textit{`throat condition'}) to develop. First, we take an embedded geometry of the metric type Eq. (\ref{intmet}) on a $t = const$, $\theta = \pi/2$ spatial slice.

\begin{equation}\label{3_d}
dl^2 = \left\lbrace a_0 + B(r)\right\rbrace^{2}g(r)^2 dr^{2} + \left\lbrace a_0 + B(r)\right\rbrace^{2}r^2 d\phi^{2}.
\end{equation}

$a_0$ is the value of $A(t)$ on the constant time slice. $dl^2$ resembles the geometry of a surface of revolution given by $\rho = \rho(z)$ rooted in a three-dimensional space. The Euclidean line element has a standard form
\begin{equation}\label{R3d}
dl^2 = dz^2 + d\rho^2 + \rho^2 d\phi^2,
\end{equation}
in cylindrically symmetric coordinates. We compare Eqs. (\ref{3_d}) and (\ref{R3d}) to write
\begin{eqnarray}\label{condd0}
&&\rho^2 = \left\lbrace a_0 + B(r)\right\rbrace^{2}r^2,\\&&\label{condd1}
dz^{2} + d\rho^{2} = \left\lbrace a_0 + B(r)\right\rbrace^{2}g(r)^2 dr^2.
\end{eqnarray}

For a constant $t$ we can further derive
\begin{equation}\label{condd2}
d\rho^2 = \left[\lbrace a_0 + B(r)\rbrace + rB'(r)\right] dr^2.
\end{equation}

Using Eqs. (\ref{condd1}) and (\ref{condd2}), some straightforward simplifactions allow us to write
\begin{eqnarray}\label{condd3}
\left(\frac{d\rho}{dz} \right)^2= \frac{\left[\lbrace a_0 + B(r)\rbrace + rB'(r)\right]^2}{\left[\lbrace (a_0 + B(r))(g(r) + 1)\rbrace + rB'(r)\right]\left[\lbrace (a_0 + B(r))(g(r) - 1)\rbrace - rB'(r)\right]}. 
\end{eqnarray}

The wormhole throat is supposed to generate a simple circle of radius $\rho$ in a diagram embedded on the surface of revolution. Around the wormhole throat a converging bundle of geodesics become parrallel. Therefore $\rho(z)$, the radius of the embedded circle must have a minima at some non-zero $r = r_w$ which implies,
\begin{equation}\label{ee4}
\frac{\left[\lbrace a_0 + B(r)\rbrace + rB'(r)\right]^2}{\left[\lbrace (a_0 + B(r))(g(r) + 1)\rbrace + rB'(r)\right]\left[\lbrace (a_0 + B(r))(g(r) - 1)\rbrace - rB'(r)\right]} \Bigg|_{r_{w}} = 0.
\end{equation}

It is a fair argument to expect that the function on the denominator is non-singular for all values of $r$. Therefore, the necessary throat condition simply becomes 
\begin{equation}\label{throatc}
\left[\lbrace a_0 + B(r)\rbrace + rB'(r)\right] \Bigg|_{r_{w}} = 0.
\end{equation}

We see that the throat condition depends on the functional form of $B(r)$ and can be satisfied for a wide range of functional form of the same. For the choice considered in this manuscript, i.e., $B(r) = \alpha e^{\lambda(r_{0}- r)}$ ($\alpha > 0$) the condition (\ref{throatc}) translates into,
\begin{equation}
\alpha e^{\lambda(r_{0}- r_{w})} = \frac{a_0}{r_{w}\lambda - 1}.
\end{equation}

We expect the value of $A(t)$ on the constant time slice, i.e., $a_0$ to be positive. Given that the left hand side of the above equation is always positive, we find a constraint on the wormhole throat radius as follows

\begin{equation}
r_{w} > \frac{1}{\lambda}.
\end{equation}

Therefore, the choice of $\lambda$ leaves a signature on the behavior of an initially converging family of geodesics and determines the wormhole throat radius. The larger we choose the value of $\lambda$, the smaller the wormhole throat radius becomes.

\section{Energy Conditions of the Exterior Geometry}
Before concluding the manuscript, we address the nature of the energy-momentum distribution of the exterior. We assume that the exterior is a generalized Vaidya geometry, 
\begin{equation}
ds^{2} = -\left[1 - \frac{2M(u, R)}{R}\right]du^{2} + 2 \epsilon dudR + r^{2} d\Omega^{2},\;\;(\epsilon = \pm 1).
\end{equation}
Here $\epsilon = \pm 1$ distinguishes between an Eddington advanced time $u$ and an Eddington retarded time $v$, respectively. The energy contained within the collapsing sphere is given by the mass function $M(u, R)$. For this metric, the $G_{\mu\nu}$ components are given by,
\begin{equation}
G^{0}_{0} = G^{1}_{1} = - \frac{2 M_R}{R^{2}} ~,~ G^{1}_{0} = \frac{2 {M_u}}{R^{2}} ~,~ G^{2}_{2} = G^{3}_{3} = - \frac{ M_{RR}}{R},
\end{equation}
where $M_u\equiv\frac{\partial M(u, R)}{\partial u}$ and $M_R\equiv\frac{\partial M(u, R)}{\partial R}$.
We write the energy momentum tensor of the exterior \cite{PhysRevD.53.R1759} as,
\begin{equation}\label{newvaidyaT}
T_{\mu\nu} = T^{(n)}_{\mu\nu} + T^{(m)}_{\mu\nu} ~,~ T^{(n)}_{\mu\nu} = \mu l_{\mu}l_{\nu} ~,~ T^{(m)}_{\mu\nu} = (\rho_\mathrm{e} + P) \left(l_{\mu}n_{\nu} + l_{\nu}n_{\mu}\right) + P g_{\mu\nu},
\end{equation}
where the two null-vectors $l_{\mu}$ and $n_{\mu}$ are usually defined as,
\begin{equation}
l_{\mu} = \delta^{0}_{\mu} ~,~ n_{\mu} = \frac{1}{2}\left[1 - \frac{2M(u, R)}{R}\right]\delta^{0}_{\mu} - \epsilon \delta^{1}_{\mu} ~,~ l_{\lambda}l^{\lambda} = n_{\lambda}n^{\lambda} = 0 ~,~ l_{\lambda}n^{\lambda} = - 1.
\end{equation}
Consequently, we can express the usual physical quantities as,
\begin{equation}
\mu = \frac{2 \epsilon {M_u}}{\kappa R^{2}} ~,~ \rho_\mathrm{e} = \frac{ 2 M_R}{\kappa R^{2}} ~,~P = - \frac{ M_{RR}}{\kappa R}.
\end{equation}

We note that $T^{(n)}_{\mu\nu}$ characterizes a flow of matter along the $u = constant$ hypersurface. Defining an orthonormal basis \cite{wangvaidya},  
\begin{equation}
E_{0}^{\mu} = \frac{l_{\mu} + n_{\mu}}{\sqrt{2}} ~,~ E_{1}^{\mu} = \frac{l_{\mu} - n_{\mu}}{\sqrt{2}} ~,~ E_{2}^{\mu} = \frac{1}{r}\delta^{\mu}_{2} ~,~ E_{3}^{\mu} = \frac{1}{r\sin\theta}\delta^{\mu}_{3},
\end{equation}
we can write a much familiar expression for the energy momentum tensor for the exterior as,
\begin{equation}
T_{ab} = \left[
\begin{array}{lccl}
\frac{\mu}{2} + \rho& \frac{\mu}{2}& 0 & 0\\
\frac{\mu}{2} & \frac{\mu}{2} - \rho & 0 & 0\\
0 & 0 & P & 0\\
0 & 0 & 0& P\\
\end{array} \right].
\end{equation}
The common classification for this class is Type-$II$ fluid. The energy condition for this fluid can be expressed in a standard manner depending on the mass function. The WEC and the SEC, for example, are written as,

\begin{equation}
\mu \geq 0, \;\;\; \rho \geq 0, \;\;\; P \geq 0,\; (\mu \neq 0).
\end{equation}
We have focused on a specific case, namely $M(u, R) = M(u)$, throughout this manuscript. Consequently, the energy conditions for the exterior reduce into a single inequality,
\begin{equation}
\mu \sim -\frac{2\frac{dM(u)}{du}}{\kappa R^{2}} \geq 0.
\end{equation}

The Misner-Sharp mass function $M(t,r)$ for the interior, evaluated using Eq. (\ref{msmf}), is given by,
\begin{equation}
M(t,r) = \frac{r(A+B)}{2} \left[1 - \frac{r^2 \dot{A}^2}{f^2 (A+B)^2} - \frac{(A+B+rB')^2}{g^2 (A+B)^2} \right].
\end{equation}

If an exterior observer immersed in the Vaidya metric can only observe the boundary hyper-surface at $r = r_{0}$, then evaluation of the above expression at this particular radial coordinate value gives the following result,
\begin{equation}
\begin{split}
 & M(t,r)_{r = r_0} \\ & = \frac{r_{0}\left\lbrace A(t) + \alpha\right\rbrace}{2} \left[1 - \frac{\beta^{2\gamma} r_{0}^{2}\dot{A}^2}{(1-\beta^{1+\gamma}) \left\lbrace A(t)+\alpha \right\rbrace^{2}} - \frac{\left\lbrace A(t) + \alpha - r_{0}\alpha\lambda \right\rbrace^{2}(1-\beta^{1+\gamma})}{\left\lbrace A(t)+\alpha \right\rbrace^{2}} \right].
\end{split}
\end{equation}

One can use this expression to evaluate and confirm that $\frac{2\frac{dM(u)}{du}}{\kappa R^{2}} \leq 0$. On the other hand, our metric construction as in Eq. (\ref{intmet}), together with the inhomogeneity profiles as in Eq. (\ref{inhomo}) clearly show that at the radial value of boundary hypersurface ($r = r_0$) the metric for $\lambda>0$ exactly represents a quantum-corrected Schwarzchild black hole (apart from a constant conformal factor). Therefore, the quantum correction will be negligible in the regions outside the spherical cluster of matter as per the original construction of \cite{PhysRevLett.121.241301, PhysRevD.98.126003, doi:10.1142/S0218271820500765} and the rest of the matching should be trivial.

\section{Conclusion}
The continuing motivation of studying gravitational collapse with a geometric formalism is many-fold. This manuscript is a consolidated attempt to address two questions simultaneously : (i) how an evolving boundary hyersurface of a massive, imploding star behaves and (ii) how the interior fluid distribution undergoes a transition from the classical into a quantum length scale.   \\

We have proposed the time-varying generalization of a solution classified as quantum-inspired black hole solution. The time evolution is always regular, comes into the metric through an inhomogeneous time-dependent conformal factor and the model describes a spherically symmetric gravitational collapse. A crucial feature of this solution is that it incorporates the mass-independent upper bound on the curvature scalars of the original static metric which comes in from the loop-quantum gravity corrections. The interior geometry carries a spherical distribution of imperfect fluid. Nature of the fluid components are not specified at the outset and treated as unknown functions in the field equations. The two metric patches involved in this process, the quantum-inspired interior and a Vaidya exterior solution are smoothly matched at the boundary hypersurface. The continuity of extrinsic curvature provides a differential equation governing the time evolution of the collapsing sphere, which is numerically solved for an extensive range of parameters to unveil the physics involved.  \\

The fate of this spherical geometry is governed by the trinity of inhomogeneities, namely $B(r) = \alpha  \exp \left[\lambda  (r_0-r)\right]$, $r_s(r) = \beta  r \exp (r-r_0)$ and $\epsilon (r) = \gamma \left[\exp (r-r_0)\right]^{-\frac{2}{3}}$. The last two functions are inspired from the original loop-quantum corrected  static solution where they are just constants. In particular, $\epsilon$ is related to the mentioned mass-independent upper bound of curvature scalars with an estimated value of $\epsilon \sim 10^{-26}$, for static solar mass black holes. In the proposed solution the form of these functions are constructed carefully such that at any instant of time the evolving geometry simply reduces into that of the original static version at $r = r_0$ (the boundary value of radial coordinate). The function $B(r)$ is also hand-crafted into the conformal factor and, as it turns out, plays a crucial role in determining the fate of the collapse. $\alpha$, $\beta$, $\gamma$ and $\lambda$ are parameters of the model. For the first three parameters the qualitative nature of the evolution (broadly defined) remains somewhat similar apart from specific scaling or generation of periodicities (generation of periodicities depends on $\alpha$). However, $\lambda$ can be thought of as a critical parameter of the model. The broad class of models characterized by $\lambda < 0$ shows an eventual collapse to singularity. On the other hand, any model with $\lambda > 0$ produces an example of a gravitational collapse ending up in a spherically symmetric, non-singular static geometry, quite similar to the loop-quantum corrected metric modulo an inhomogeneous conformal factor. \\

The behavior of the interior fluid distribution is studied in details for $\lambda = \pm 1$. The collapsing shells obey standard energy conditions broadly, but not always, and is heavily dependent on the nature of evolution and the regions we are looking at. For instance, the high matter density near the collapsing core overpowers the effective energy density due to perturbative quantum effect (which is negative). As a consequence WEC is satisfied. However, near the boundary hyprsurface where there is less matter accummulation WEC is clearly violated. On the other hand, the SEC and NEC are violated for a much larger portion of the interior, when the final solution is non-singular ($\lambda > 0$) compared to the case when a singularity is formed ($\lambda < 0$). Such a violation of standard energy conditions is clearly associated with the quantum corrections affecting the clustered matter distribution. The dominance of the quantum corrections can be seen in two phases. For $\lambda < 0$ case, there exist situations when the quantum effects are able to cause a bounce but are not strong enough to prevent the ultimate formation of the singularity. On the other hand, for  $\lambda > 0$ case, the quantum corrections indeed prevent the final crunch and produce a non-singular static configuration.  \\ 

We have also proved that the singularity, formed in the $\lambda < 0$ case, is always hidden inside an apparent horizon (trapped surface). Therefore, the central singularity in this case is a black hole singularity. The system is found to contain three horizons at some intermediate stage during the collapse. However, finally a single horizon exists which envelope the singularity. On the other hand, for the case of $\lambda > 0$, there is a possibility that the sphere evolves into a wormhole throat. This is checked by studying an embedded geometry of the metric on a constant-time spatial slice. The wormhole throat radius, intriguingly, depends on the value of $\lambda$, which is a tributary of the inhomogeneity in the conformal factor. It is easy to connect this parameter with the four-velocity of the comoving observer ($\sim \frac{1}{\sqrt{-g^{00}}}$), as well as with the initial volume of the sphere (proportional to the conformal factor). We conclude with an introspection that, these parameters might have a more fundamental connection with horizon entropy and the thermodynamic behavior of the resulting black-hole/black-hole mimickers. This has never been investigated in the past and shall be included in a subsequent manuscript.

\bibliographystyle{JHEP}
\bibliography{ref}

\providecommand{\href}[2]{#2}\begingroup\raggedright\begin{thebibliography}{10}

\bibitem{Datt1938}
B.~Datt, \emph{{\"U}ber eine klasse von l{\"o}sungen der
  gravitationsgleichungen der relativit{\"a}t},
  \href{https://doi.org/10.1007/BF01374951}{\emph{Zeitschrift f{\"u}r Physik}
  {\bfseries 108} (1938) 314}.

\bibitem{PhysRev.55.374}
J.R.~Oppenheimer and G.M.~Volkoff, \emph{On massive neutron cores},
  \href{https://doi.org/10.1103/PhysRev.55.374}{\emph{Phys. Rev.} {\bfseries
  55} (1939) 374}.

\bibitem{PhysRev.56.455}
J.R.~Oppenheimer and H.~Snyder, \emph{On continued gravitational contraction},
  \href{https://doi.org/10.1103/PhysRev.56.455}{\emph{Phys. Rev.} {\bfseries
  56} (1939) 455}.

\bibitem{Malafarina2016}
D.~Malafarina, \emph{A brief review of relativistic gravitational collapse},
  in \emph{Astrophysics of Black Holes: From Fundamental Aspects to Latest
  Developments}, C.~Bambi, ed., (Berlin, Heidelberg), pp.~169--197, Springer
  Berlin Heidelberg (2016),
  \href{https://doi.org/10.1007/978-3-662-52859-4_5}{DOI}.

\bibitem{Joshi2000}
P.S.~Joshi, \emph{Gravitational collapse: The story so far},
  \href{https://doi.org/10.1007/s12043-000-0164-4}{\emph{Pramana} {\bfseries
  55} (2000) 529}.

\bibitem{Yodzis1973}
P.~Yodzis, H.-J.~Seifert and H.~M{\"u}ller~zum Hagen, \emph{On the occurrence
  of naked singularities in general relativity},
  \href{https://doi.org/10.1007/BF01646443}{\emph{Communications in
  Mathematical Physics} {\bfseries 34} (1973) 135}.

\bibitem{MllerzumHagen1974}
H.~M{\"u}ller~zum Hagen, P.~Yodzis and H.-J.~Seifert, \emph{On the occurrence
  of naked singularities in general relativity. ii},
  \href{https://doi.org/10.1007/BF01646032}{\emph{Communications in
  Mathematical Physics} {\bfseries 37} (1974) 29}.

\bibitem{PhysRevD.65.101501}
P.S.~Joshi, N.~Dadhich and R.~Maartens, \emph{Why do naked singularities form
  in gravitational collapse?},
  \href{https://doi.org/10.1103/PhysRevD.65.101501}{\emph{Phys. Rev. D}
  {\bfseries 65} (2002) 101501}.

\bibitem{PhysRevD.70.087502}
P.S.~Joshi, R.~Goswami and N.~Dadhich, \emph{Why do naked singularities form in
  gravitational collapse? ii},
  \href{https://doi.org/10.1103/PhysRevD.70.087502}{\emph{Phys. Rev. D}
  {\bfseries 70} (2004) 087502}.

\bibitem{doi:10.1142/S0217732300000992}
P.S.~JOSHI, N.K.~DADHICH and R.~MAARTENS, \emph{Gamma-ray bursts as the
  birth-cries of black holes},
  \href{https://doi.org/10.1142/S0217732300000992}{\emph{Modern Physics Letters
  A} {\bfseries 15} (2000) 991}
  [\href{https://arxiv.org/abs/https://doi.org/10.1142/S0217732300000992}{{\ttfamily
  https://doi.org/10.1142/S0217732300000992}}].

\bibitem{bronni}
K.A.~Bronnikov and M.A.~Kovalchuk, \emph{Some exact models for nonspherical
  collapse?},
  \href{https://doi.org/https://doi.org/10.1007/BF00764015}{\emph{General
  Relativity and Gravitation} {\bfseries 16} (1984) 31}.

\bibitem{Penrose2002}
R.~Penrose, \emph{``golden oldie'': Gravitational collapse: The role of general
  relativity}, \href{https://doi.org/10.1023/A:1016578408204}{\emph{General
  Relativity and Gravitation} {\bfseries 34} (2002) 1141}.

\bibitem{chakra}
S.~Chakrabarti, \emph{Can a variation of fine structure constant influence the
  fate of gravitational collapse?},
  \href{https://doi.org/10.1140/epjc/s10052-023-11877-1}{\emph{Eur. Phys. J. C}
  {\bfseries 83} (2023) 693}.

\bibitem{PhysRevLett.70.9}
M.W.~Choptuik, \emph{Universality and scaling in gravitational collapse of a
  massless scalar field},
  \href{https://doi.org/10.1103/PhysRevLett.70.9}{\emph{Phys. Rev. Lett.}
  {\bfseries 70} (1993) 9}.

\bibitem{Gundlach2007}
C.~Gundlach and J.M.~Mart{\'i}n-Garc{\'i}a, \emph{Critical phenomena in
  gravitational collapse},
  \href{https://doi.org/10.12942/lrr-2007-5}{\emph{Living Reviews in
  Relativity} {\bfseries 10} (2007) 5}.

\bibitem{PhysRevD.104.024071}
S.~Chakrabarti and S.~Kar, \emph{Wormhole geometry from gravitational
  collapse}, \href{https://doi.org/10.1103/PhysRevD.104.024071}{\emph{Phys.
  Rev. D} {\bfseries 104} (2021) 024071}.

\bibitem{PhysRevD.101.124044}
S.~Chakrabarti and J.L.~Said, \emph{Geodesic congruences and a collapsing
  stellar distribution in $f(t)$ theories},
  \href{https://doi.org/10.1103/PhysRevD.101.124044}{\emph{Phys. Rev. D}
  {\bfseries 101} (2020) 124044}.

\bibitem{Chakrabarti2018}
S.~Chakrabarti, R.~Goswami, S.~Maharaj and N.~Banerjee, \emph{Conformally flat
  collapsing stars in f(r) gravity},
  \href{https://doi.org/10.1007/s10714-018-2472-3}{\emph{General Relativity and
  Gravitation} {\bfseries 50} (2018) 148}.

\bibitem{Chakrabarti2018egb}
S.~Chakrabarti, \emph{Collapsing spherical star in scalar-einstein-gauss-bonnet
  gravity with a quadratic coupling},
  \href{https://doi.org/10.1140/epjc/s10052-018-5798-9}{\emph{The European
  Physical Journal C} {\bfseries 78} (2018) 296}.

\bibitem{santosbm}
N.O.~Santos, \emph{{Non-adiabatic radiating collapse}},
  \href{https://doi.org/10.1093/mnras/216.2.403}{\emph{Monthly Notices of the
  Royal Astronomical Society} {\bfseries 216} (1985) 403}.

\bibitem{Kolassis1988}
C.A.~Kolassis, N.O.~Santos and D.~Tsoubelis, \emph{Friedmann-like collapsing
  model of a radiating sphere with heat flow},
  \href{https://doi.org/10.1086/166233}{\emph{The Astrophysical Journal}
  {\bfseries 327} (1988) 755}.

\bibitem{KOLASSIS1989243}
C.~Kolassis and N.O.~Santos, \emph{Gravitational collapse of a radiating star},
  \href{https://doi.org/https://doi.org/10.1016/0375-9601(89)90478-7}{\emph{Physics
  Letters A} {\bfseries 141} (1989) 243}.

\bibitem{oliveira}
A.K.G.~de~Oliveira, C.A.~Kolassis and N.O.~Santos, \emph{{Collapse of a
  radiating star revisited}},
  \href{https://doi.org/10.1093/mnras/231.4.1011}{\emph{Monthly Notices of the
  Royal Astronomical Society} {\bfseries 231} (1988) 1011}.

\bibitem{PhysRevD.45.2732}
F.~Fayos, X.~Ja\'en, E.~Llanta and J.M.M.~Senovilla, \emph{Interiors of
  vaidya's radiating metric: Gravitational collapse},
  \href{https://doi.org/10.1103/PhysRevD.45.2732}{\emph{Phys. Rev. D}
  {\bfseries 45} (1992) 2732}.

\bibitem{doi:10.1142/S0218271805006584}
S.D.~MAHARAJ and M.~GOVENDER, \emph{Radiating collapse with vanishing weyl
  stresses},
  \href{https://doi.org/10.1142/S0218271805006584}{\emph{International Journal
  of Modern Physics D} {\bfseries 14} (2005) 667}
  [\href{https://arxiv.org/abs/https://doi.org/10.1142/S0218271805006584}{{\ttfamily
  https://doi.org/10.1142/S0218271805006584}}].

\bibitem{PhysRevLett.96.031103}
S.A.~Hayward, \emph{Formation and evaporation of nonsingular black holes},
  \href{https://doi.org/10.1103/PhysRevLett.96.031103}{\emph{Phys. Rev. Lett.}
  {\bfseries 96} (2006) 031103}.

\bibitem{PhysRevD.94.104056}
V.P.~Frolov, \emph{Notes on nonsingular models of black holes},
  \href{https://doi.org/10.1103/PhysRevD.94.104056}{\emph{Phys. Rev. D}
  {\bfseries 94} (2016) 104056}.

\bibitem{DeLorenzo2015}
T.~De~Lorenzo, C.~Pacilio, C.~Rovelli and S.~Speziale, \emph{On the effective
  metric of a planck star},
  \href{https://doi.org/10.1007/s10714-015-1882-8}{\emph{General Relativity and
  Gravitation} {\bfseries 47} (2015) 41}.

\bibitem{PhysRevD.96.044010}
R.~Casadio, A.~Giugno, A.~Giusti and M.~Lenzi, \emph{Quantum corpuscular
  corrections to the newtonian potential},
  \href{https://doi.org/10.1103/PhysRevD.96.044010}{\emph{Phys. Rev. D}
  {\bfseries 96} (2017) 044010}.

\bibitem{PhysRevD.98.104016}
R.~Casadio, M.~Lenzi and O.~Micu, \emph{Bootstrapping newtonian gravity},
  \href{https://doi.org/10.1103/PhysRevD.98.104016}{\emph{Phys. Rev. D}
  {\bfseries 98} (2018) 104016}.

\bibitem{PhysRevLett.121.241301}
A.~Ashtekar, J.~Olmedo and P.~Singh, \emph{Quantum transfiguration of kruskal
  black holes},
  \href{https://doi.org/10.1103/PhysRevLett.121.241301}{\emph{Phys. Rev. Lett.}
  {\bfseries 121} (2018) 241301}.

\bibitem{PhysRevD.98.126003}
A.~Ashtekar, J.~Olmedo and P.~Singh, \emph{Quantum extension of the kruskal
  spacetime}, \href{https://doi.org/10.1103/PhysRevD.98.126003}{\emph{Phys.
  Rev. D} {\bfseries 98} (2018) 126003}.

\bibitem{doi:10.1142/S0218271820500765}
A.~Ashtekar and J.~Olmedo, \emph{Properties of a recent quantum extension of
  the kruskal geometry},
  \href{https://doi.org/10.1142/S0218271820500765}{\emph{International Journal
  of Modern Physics D} {\bfseries 29} (2020) 2050076}
  [\href{https://arxiv.org/abs/https://doi.org/10.1142/S0218271820500765}{{\ttfamily
  https://doi.org/10.1142/S0218271820500765}}].

\bibitem{PhysRevD.76.104030}
C.G.~B\"ohmer and K.~Vandersloot, \emph{Loop quantum dynamics of the
  schwarzschild interior},
  \href{https://doi.org/10.1103/PhysRevD.76.104030}{\emph{Phys. Rev. D}
  {\bfseries 76} (2007) 104030}.

\bibitem{PhysRevD.78.064040}
D.-W.~Chiou, \emph{Phenomenological loop quantum geometry of the schwarzschild
  black hole}, \href{https://doi.org/10.1103/PhysRevD.78.064040}{\emph{Phys.
  Rev. D} {\bfseries 78} (2008) 064040}.

\bibitem{Corichi_2016}
A.~Corichi and P.~Singh, \emph{Loop quantization of the schwarzschild interior
  revisited},
  \href{https://doi.org/10.1088/0264-9381/33/5/055006}{\emph{Classical and
  Quantum Gravity} {\bfseries 33} (2016) 055006}.

\bibitem{Olmedo_2017}
J.~Olmedo, S.~Saini and P.~Singh, \emph{From black holes to white holes: a
  quantum gravitational, symmetric bounce},
  \href{https://doi.org/10.1088/1361-6382/aa8da8}{\emph{Classical and Quantum
  Gravity} {\bfseries 34} (2017) 225011}.

\bibitem{BOUHMADILOPEZ2020100701}
M.~Bouhmadi-López, S.~Brahma, C.-Y.~Chen, P.~Chen and D.~han Yeom,
  \emph{Asymptotic non-flatness of an effective black hole model based on loop
  quantum gravity},
  \href{https://doi.org/https://doi.org/10.1016/j.dark.2020.100701}{\emph{Physics
  of the Dark Universe} {\bfseries 30} (2020) 100701}.

\bibitem{chanbm}
R.~Chan, \emph{{Radiating gravitational collapse with shear viscosity}},
  \href{https://doi.org/10.1046/j.1365-8711.2000.03547.x}{\emph{Monthly Notices
  of the Royal Astronomical Society} {\bfseries 316} (2000) 588}.

\bibitem{sym12081264}
V.~Faraoni and A.~Giusti, \emph{Unsettling physics in the quantum-corrected
  schwarzschild black hole},
  \href{https://doi.org/10.3390/sym12081264}{\emph{Symmetry} {\bfseries 12}
  (2020) }.

\bibitem{Israel1967}
W.~Israel, \emph{Singular hypersurfaces and thin shells in general relativity},
  \href{https://doi.org/10.1007/BF02712210}{\emph{Il Nuovo Cimento B
  (1965-1970)} {\bfseries 48} (1967) 463}.

\bibitem{10.1046/j.1365-8711.2000.03547.x}
R.~Chan, \emph{{Radiating gravitational collapse with shear viscosity}},
  \href{https://doi.org/10.1046/j.1365-8711.2000.03547.x}{\emph{Monthly Notices
  of the Royal Astronomical Society} {\bfseries 316} (2000) 588}
  [\href{https://arxiv.org/abs/https://academic.oup.com/mnras/article-pdf/316/3/588/2918056/316-3-588.pdf}{{\ttfamily
  https://academic.oup.com/mnras/article-pdf/316/3/588/2918056/316-3-588.pdf}}].

\bibitem{PhysRev.136.B571}
C.W.~Misner and D.H.~Sharp, \emph{Relativistic equations for adiabatic,
  spherically symmetric gravitational collapse},
  \href{https://doi.org/10.1103/PhysRev.136.B571}{\emph{Phys. Rev.} {\bfseries
  136} (1964) B571}.

\bibitem{1966ApJhm}
J.~Hernandez, Walter~C. and C.W.~Misner, \emph{{Observer Time as a Coordinate
  in Relativistic Spherical Hydrodynamics}},
  \href{https://doi.org/10.1086/148525}{\emph{The Astrophysical Journal}
  {\bfseries 143} (1966) 452}.

\bibitem{10.1063/1.1665273}
M.E.~Cahill and G.C.~McVittie, \emph{{Spherical Symmetry and Mass‐Energy in
  General Relativity. I. General Theory}},
  \href{https://doi.org/10.1063/1.1665273}{\emph{Journal of Mathematical
  Physics} {\bfseries 11} (2003) 1382}
  [\href{https://arxiv.org/abs/https://pubs.aip.org/aip/jmp/article-pdf/11/4/1382/11147781/1382\_1\_online.pdf}{{\ttfamily
  https://pubs.aip.org/aip/jmp/article-pdf/11/4/1382/11147781/1382\_1\_online.pdf}}].

\bibitem{Saghafi2023}
S.~Saghafi and K.~Nozari, \emph{Hawking-like radiation as tunneling from a
  cosmological black hole in modified gravity: semiclassical approximation and
  beyond}, \href{https://doi.org/10.1007/s10714-022-03063-7}{\emph{General
  Relativity and Gravitation} {\bfseries 55} (2023) 20}.

\bibitem{PhysRevLett.128.121301}
V.~Husain, J.G.~Kelly, R.~Santacruz and E.~Wilson-Ewing, \emph{Quantum gravity
  of dust collapse: Shock waves from black holes},
  \href{https://doi.org/10.1103/PhysRevLett.128.121301}{\emph{Phys. Rev. Lett.}
  {\bfseries 128} (2022) 121301}.

\bibitem{PhysRevD.106.024014}
V.~Husain, J.G.~Kelly, R.~Santacruz and E.~Wilson-Ewing, \emph{Fate of quantum
  black holes}, \href{https://doi.org/10.1103/PhysRevD.106.024014}{\emph{Phys.
  Rev. D} {\bfseries 106} (2022) 024014}.

\bibitem{PhysRevD.106.046012}
S.~Hergott, V.~Husain and S.~Rastgoo, \emph{Model metrics for quantum black
  hole evolution: Gravitational collapse, singularity resolution, and transient
  horizons}, \href{https://doi.org/10.1103/PhysRevD.106.046012}{\emph{Phys.
  Rev. D} {\bfseries 106} (2022) 046012}.

\bibitem{BARCELO1999127}
C.~Barceló and M.~Visser, \emph{Traversable wormholes from massless
  conformally coupled scalar fields},
  \href{https://doi.org/https://doi.org/10.1016/S0370-2693(99)01117-X}{\emph{Physics
  Letters B} {\bfseries 466} (1999) 127}.

\bibitem{PhysRevD.54.6233}
E.E.~Flanagan and R.M.~Wald, \emph{Does back reaction enforce the averaged null
  energy condition in semiclassical gravity?},
  \href{https://doi.org/10.1103/PhysRevD.54.6233}{\emph{Phys. Rev. D}
  {\bfseries 54} (1996) 6233}.

\bibitem{PhysRevD.53.R1759}
V.~Husain, \emph{Exact solutions for null fluid collapse},
  \href{https://doi.org/10.1103/PhysRevD.53.R1759}{\emph{Phys. Rev. D}
  {\bfseries 53} (1996) R1759}.

\bibitem{wangvaidya}
A.~Wang and Y.~Wu, \emph{Letter: Generalized vaidya solutions},
  \href{https://doi.org/10.1103/PhysRevD.53.R1759}{\emph{General Relativity and
  Gravitation} {\bfseries 31} (1999) 107}.

\end{thebibliography}\endgroup

\end{document}